\documentclass[smallextended]{svjour3}       

\usepackage[utf8]{inputenc}
\usepackage{algorithmic}
\usepackage{multirow}
\usepackage[linesnumbered]{algorithm2e}
\usepackage{capt-of}
\usepackage{amsmath,tabularx}
\usepackage[table]{xcolor}
\usepackage{graphics}
\usepackage{graphicx}
\usepackage[skip=4pt,compatibility=false]{caption}
\usepackage{subcaption}
\usepackage{listings}
\usepackage{xcolor}
\usepackage{proof}
\usepackage{stmaryrd}
\usepackage{xspace}
\usepackage{amssymb}
\usepackage{epstopdf}
\usepackage{tabularx}
\usepackage{rotating}										
\usepackage{booktabs} 

\usepackage{balance}
\usepackage{subscript}
\usepackage{url}
\usepackage{makecell}

\newcommand{\pluseq}{\mathrel{+}=}

\newcommand{\asgn}{\mathrel{:}=}
\newcommand{\prmt}{\textcolor{blue}{(\toolname) }}

\newcommand{\alg}[1]{Algorithm~\ref{#1}}
\newcommand{\fig}[1]{Figure~\ref{#1}}
\newcommand{\tab}[1]{Table~\ref{#1}}
\newcommand{\sect}[1]{\S\ref{#1}}
 
\definecolor{codegreen}{rgb}{0,0.6,0}
\definecolor{codegray}{rgb}{0.5,0.5,0.5}
\definecolor{codepurple}{rgb}{0.58,0,0.82}
\definecolor{backcolour}{rgb}{0.95,0.95,0.92}
\definecolor{aliceblue}{rgb}{0.94,0.97,0.80}

\newcolumntype{C}{>{\centering\arraybackslash}X}%

\newcommand{\B}[1]{\langle #1 \rangle}
\definecolor{arylideyellow}{rgb}{0.91,0.84,0.42}

\lstdefinestyle{mystyle}{
    backgroundcolor=\color{backcolour},   
    commentstyle=\color{codegreen},
    keywordstyle=\color{magenta},
    numberstyle=\tiny\color{codegray},
    stringstyle=\color{codepurple},
    breakatwhitespace=false,         
    breaklines=true,                 
    captionpos=b,                    
    keepspaces=true,                 
    numbers=left, 
    xleftmargin=6pt,                   
    numbersep=2pt,                  
    showspaces=false,                
    showstringspaces=false,
    showtabs=false,                  
    tabsize=2,
    mathescape=true,
    escapeinside=`'
}
 
\newcommand{\toolname}{\textsc{Wolverine2}\xspace}
\newcommand{\oldtoolname}{\textsc{Wolverine}\xspace}
\newcommand{\porg}{\mathcal{P}\xspace} 
\newcommand{\prep}{\widehat{\mathcal{P}}\xspace} 
\newcommand{\st}{\mathcal{S}\xspace} 
\newcommand{\var}[1]{\texttt{#1}} 

\newcommand{\fontsmall}{\fontsize{8pt}{8pt}\selectfont}
\newcommand{\colbox}[1]{\vspace{2pt}{\noindent{\colorbox{aliceblue}{\begin{minipage}{0.95\columnwidth} \fontsmall{\texttt{#1}} \end{minipage}}}}\vspace{2pt}}

\newcommand{\opsemf}[3]{(#1, #2) \overset{\Delta}{\implies} #3}
\newcommand{\opsemb}[3]{(#1, #2) \overset{\nabla}{\implies} #3}

\newcommand{\hoarerule}[3]{\text{\{}#1\text{\}} \ #2 \ \text{\{}#3\text{\}}}

\let\subparagraph\relax

\usepackage{enumitem}
\setitemize{noitemsep,topsep=0pt,parsep=0pt,partopsep=0pt}

\begin{document}

\title{Debug-Localize-Repair: A Symbiotic Construction for Heap Manipulations}

\author{Sahil Verma         \and
        Subhajit Roy 
}


\institute{
        Sahil Verma \at
        Paul G. Allen School of Computer Science and Engineering, \\ University of Washington, Seattle, USA \\
        \email{vsahil@cs.washington.edu}
        \and
        Subhajit Roy \at
        Department of Computer Science and Engineering, \\ Indian Institute of Technology Kanpur, India \\
        \email{subhajit@cse.iitk.ac.in}
}

\date{Received: date / Accepted: date}

\begin{abstract}
{\small
        We present \toolname, an integrated Debug-Localize-Repair environment for heap manipulating programs. \toolname provides an interactive debugging environment: while concretely executing a program via on an interactive shell supporting common debugging facilities, \toolname displays the abstract program states (as box-and-arrow diagrams) as a visual aid to the programmer, packages a novel, proof-directed repair algorithm to quickly synthesize the repair patches and a new bug localization algorithm to reduce the search space of repairs. \toolname supports ``hot-patching" of the generated patches to provide a seamless debugging environment, and also facilitates new debug-localize-repair possibilities: \textit{specification refinement} and \textit{checkpoint-based hopping}.

        We evaluate \toolname on 6400 buggy programs (generated using automated fault injection) on a variety of data-structures like singly, doubly, and circular linked lists, AVL trees, Red-Black trees, Splay Trees and Binary Search Trees; 
        \toolname could repair all the buggy instances within realistic programmer wait-time (less than 5 sec in most cases). 
        \toolname could also repair more than 80\% of the 247 (buggy) student submissions where a reasonable attempt was made.
}
\end{abstract}


\maketitle


\section{Introduction}
\label{sec:introduction}
Hunting for bugs in a heap manipulating program is a hard proposition. 
We present \toolname, an integrated debugging-localize-repair tool for heap-manipulating programs.
\toolname uses \var{gdb}~\cite{gdb} to control the concrete execution of the buggy program to provide a live visualization of the program (abstract) states as box-and-arrow diagrams. 
Programmers routinely use such box-and-arrow diagrams to plan heap manipulations and in online education~\cite{Guo:2013}.

Similar to popular debugging tools, \toolname packages common debugging facilities like stepping through an execution, setting breakpoints, fast-forwarding to a breakpoint (see \tab{tab:cheatsheet}). 
At the same time, \toolname provides additional commands for driving in situ repair: whenever the programmer detects an unexpected program state or control-flow (indicating a buggy execution), she can {\it repair} the box-and-arrow diagram to the expected state or force the expected control-flow (like forcing another execution of a \var{while} loop though the loop-exit condition is satisfied) during the debugging session. 
These expectations from the programmer are captured by \toolname as constraints to build a (partial) specification.

\begin{table}[t]
\centering
\caption{\label{tab:cheatsheet}\toolname cheatsheet}
\scalebox{0.9}{
\begin{tabular}{| c | c |}
\hline
Command & Action\\
\hline
\texttt{start} & Starts execution \\
\texttt{enter}, \texttt{leave} & Enter/exit loop\\
\texttt{next} & Executes next statement\\
\texttt{step} & Step into a function\\
\texttt{change} $e_s$ $v_s$ & Set entity $e_s$ to value $v_s$\\
\texttt{spec} & Add program state to specification\\
\texttt{repair} & Return repaired code\\
\texttt{rewrite} & Rewrite the patched file as a C program\\
\hline
\end{tabular}
}
\end{table}

When the programmer feels that she has communicated enough constraints to the tool, she can issue a \var{repair} command, requesting \toolname to attempt an automated repair. 
\toolname is capable of simulating {\it hot-patching} of the repair patch (generated by its repair module), allowing the debugging session to continue from the same point without requiring the user to abort the debug session, recompile the program with the new repair patch and start debugging. 
As the repair patch is guaranteed to have met all the user expectations till this point, the programmer can seamlessly continue the debugging session from the same program point, with the repair-patch applied, without requiring an abort-compile-debug cycle. This debug-repair scheme requires the user to point out the faults in the program states, while \toolname takes care of correcting (repairing) the fault in the underlying program.


\toolname enables a seamless integration of debugging, fault-localization and repair (\textit{debug-localize-repair}), thereby facilitates novel debug strategies wherein a skilled developer can drive faster repairs by communicating her domain knowledge to \toolname: if the programmer has confidence that a set of statements cannot have a bug, she can use {\it specification refinement} to eliminate these statements from the repair search space. 
Hence, rather than eliminating human expertise, \toolname allows a synergistic human-machine interaction. Additionally, \toolname allows for a new repair-space exploration strategy, that we refer to as \textit{checkpoint-based hopping}\footnote{We thank the anonymous reviewers of the preliminary conference version of this paper for suggesting this feature.}, wherein the developer can explore multiple strategies of fixing the program simultaneously, examine the repairs along each direction, and switch between the different candidate fixes seamlessly---to converge to the final fix.

\toolname bundles a novel \textit{proof-directed repair strategy}: it generates a repair constraint that underapproximates the potential repair search space (via additional \textit{underapproximation constraints}). If the repair constraint is satisfiable, a repair patch is generated. If proof of unsatisfiability is found (indicating a failed repair attempt) that does not depend on an underapproximation constraint, it indicates a buggy specification or a structural limitation in the tool's settings; else, the respective underapproximation constraint that appears in the proof indicates the widening direction. 

To further improve the scalability of repair, we also design an inexpensive bug localization technique that identifies suspicious statements by tracking the difference in the states in the \textit{forward execution} (proceeding from the precondition to the postcondition) and an (abstract) \textit{backward execution} (commencing from the postcondition to the precondition). The algorithm leverages on an insightful result that buggy statements always appear at program locations that exhibit a {\it non-zero gradient on the state differences} between the forward and backward execution. We prove that our algorithm is sound, i.e., it overapproximates the set of faulty statements, thereby shrinking the repair space appreciably without missing out on the ground truth bug. Our experiments show that this algorithm can shrink the suspicious statements to less than 12\% of the program size in 90\% of our benchmarks and works better than popular statistical bug localization techniques.

We evaluate \toolname on a set of 6400 buggy files: 40~randomly generated faulty versions over four faulty configurations of 40 benchmark programs collected from online sources~\cite{GeeksForGeeks} spanning multiple data-structures like singly, doubly and circular linked lists, Binary Search Trees, AVL trees, Red-Black trees, and Splay trees. We classify the 40 programs into two categories:\textit{smaller} (20 programs) and larger (20 programs) based on the program size. \toolname successfully repairs all faults in the benchmarks within a reasonable time (less than 5 seconds for most programs). 
To evaluate the effectiveness of our bug localization algorithm, we switch off bug localization before repair: \toolname slows down by more than 225$\times$ without bug localization on our larger benchmarks and fails to repair 1262 programs (out of 6400) within a timeout of 300s.

We also evaluate \toolname on 247 student submissions from an introductory programming course~\cite{Prutor16}, consisting of problems for five heap manipulating problems; \toolname could repair more than 80\% of the programs where the student had made a reasonable attempt.

\noindent We make the following contributions in this paper:
\begin{itemize}[noitemsep,wide=5pt, leftmargin=20pt]
 \item We propose that an integrated debug-localize-repair environment can yield significant benefits; we demonstrate it by building a tool, \toolname, to facilitate debug-localize-repair on heap manipulations;
 \item We propose a new \textit{proof-directed repair strategy} that uses the proof of unsatisfiability to guide the repair along the most promising direction;
 \item We propose advanced debugging techniques, \textit{specification refinement} and \textit{checkpoint-based hopping}, that are facilitated by this integration of debugging and repair.
 \item We design a new fault localization algorithm for heap manipulating programs based on the gradient between the states in a forward and backward execution. 
\end{itemize}

\toolname extends our previous work on \oldtoolname~\cite{Verma:2017}: \toolname augments the abilities of \oldtoolname with a new module for bug localization (\sect{sec:localization}), which has significantly improved (33-779$\times$) its runtime performance, allowing it to solve many instances that were beyond \oldtoolname. 
We evaluate \toolname on a larger benchmark set to demonstrate the advanced capabilities of the tool (\sect{sec:experiments}). 
We have also added new debugging capabilities (\sect{sec:checkpoint-hop}) (some of which were suggested by the reviewers of the conference version).

\section{Overview}
\label{sec:overview}
\subsection{A \toolname Debug-Localize-Repair Session}
\begin{figure}
\begin{minipage}{\textwidth}
\begin{lstlisting}[language = C]
struct node *head;
void reverse(){
    struct node *current, *temp1 = NULL, *temp2 = NULL;
    current = head;
    while (temp1 != NULL){ // FIX1: current != NULL
        temp1 = current->prev;
        temp2 = current->prev; // FIX2: current->next 
        current->prev = temp2;
        current->next = temp1; 
        current = current->prev; 
    }
    // head = temp1->prev; // FIX3: insert stmt.
}
...
int main(){
    push(2); push(4); push(8); push(10);
    reverse();
}
\end{lstlisting}
\end{minipage}
\caption{Our motivating example}
\label{fig:motivating}
\end{figure}

\begin{figure*}
 \centering
 \begin{subfigure}[b]{0.5\textwidth}
 \centering
 \includegraphics[width=\textwidth]{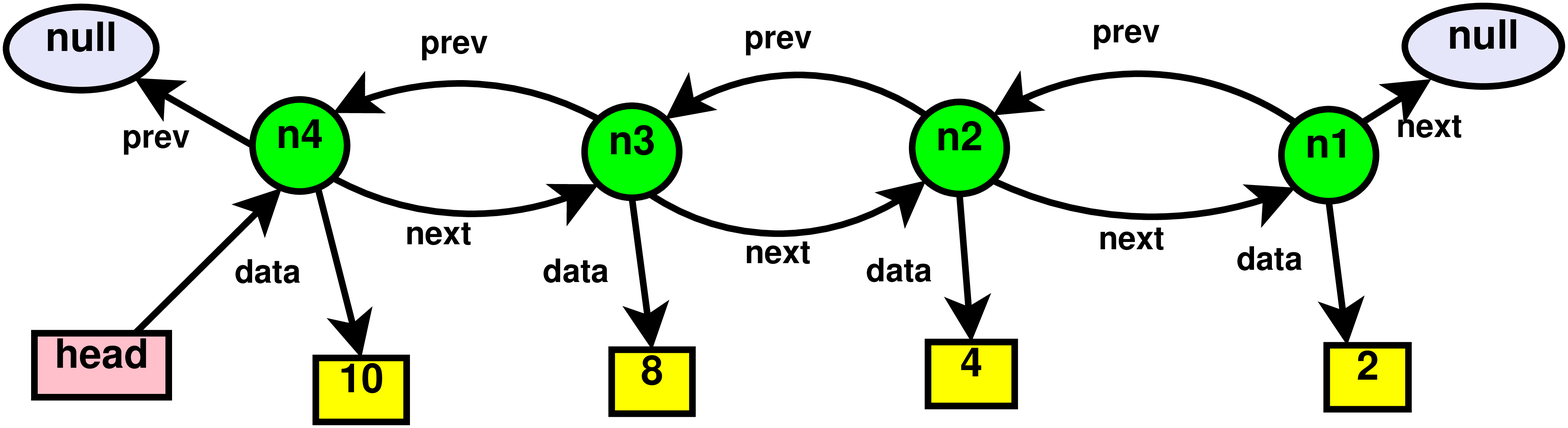}
 \caption{Nodes added (before entering reverse)}
 \label{fig:start}
 \end{subfigure}%
 \hfill
 \begin{subfigure}[b]{0.5\textwidth}
 \centering
 \includegraphics[width=\textwidth]{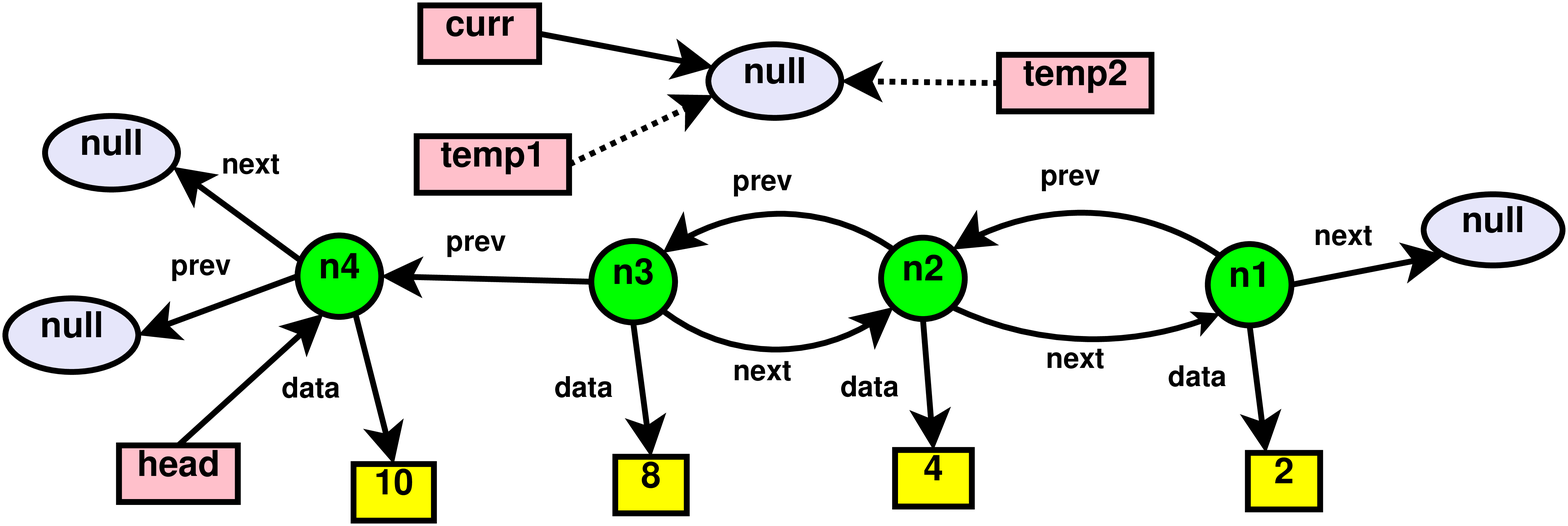}
 \caption{At the end of first loop iteration (without user changes)}
 \label{fig:first}
 \end{subfigure}%
 \\
 \begin{subfigure}[b]{0.5\textwidth}
 \centering
 \includegraphics[width=\textwidth]{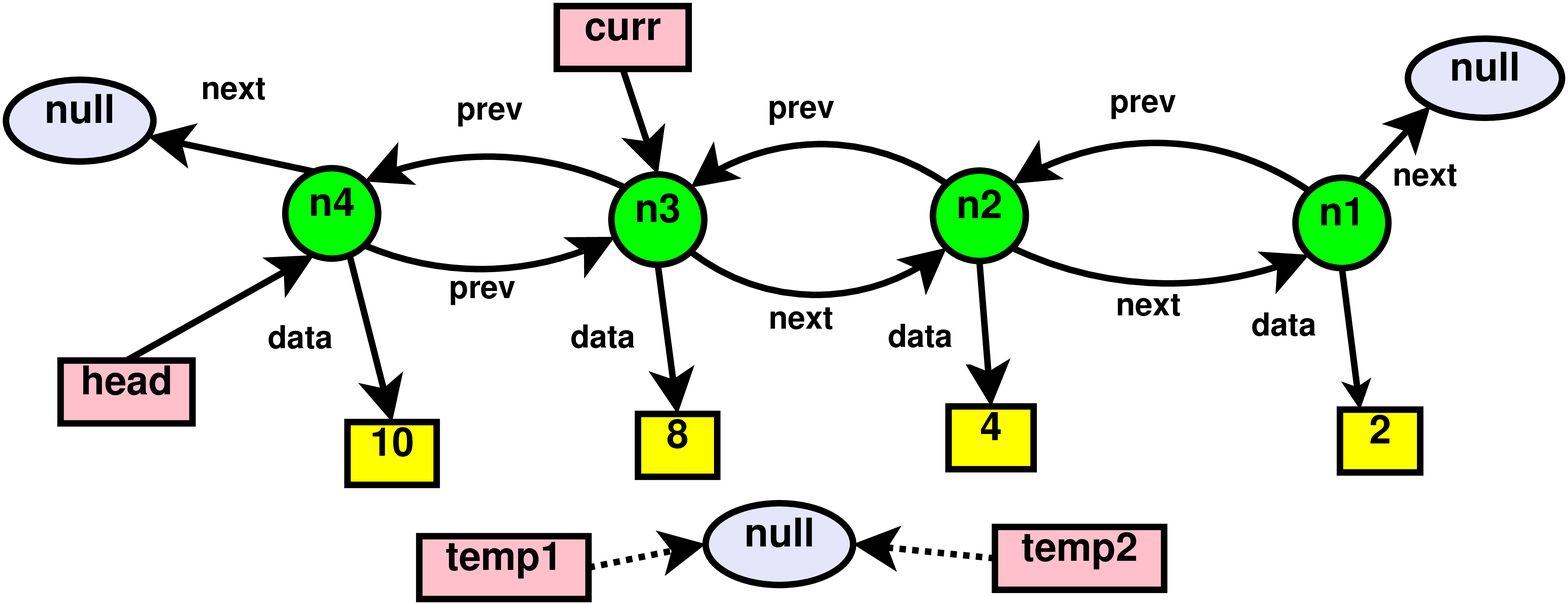}
 \caption{Changes brought about by the user after first loop iteration}
 \label{fig:change}
 \end{subfigure}%
 \hfill
 \begin{subfigure}[b]{0.5\textwidth}
 \centering
 \includegraphics[width=\textwidth]{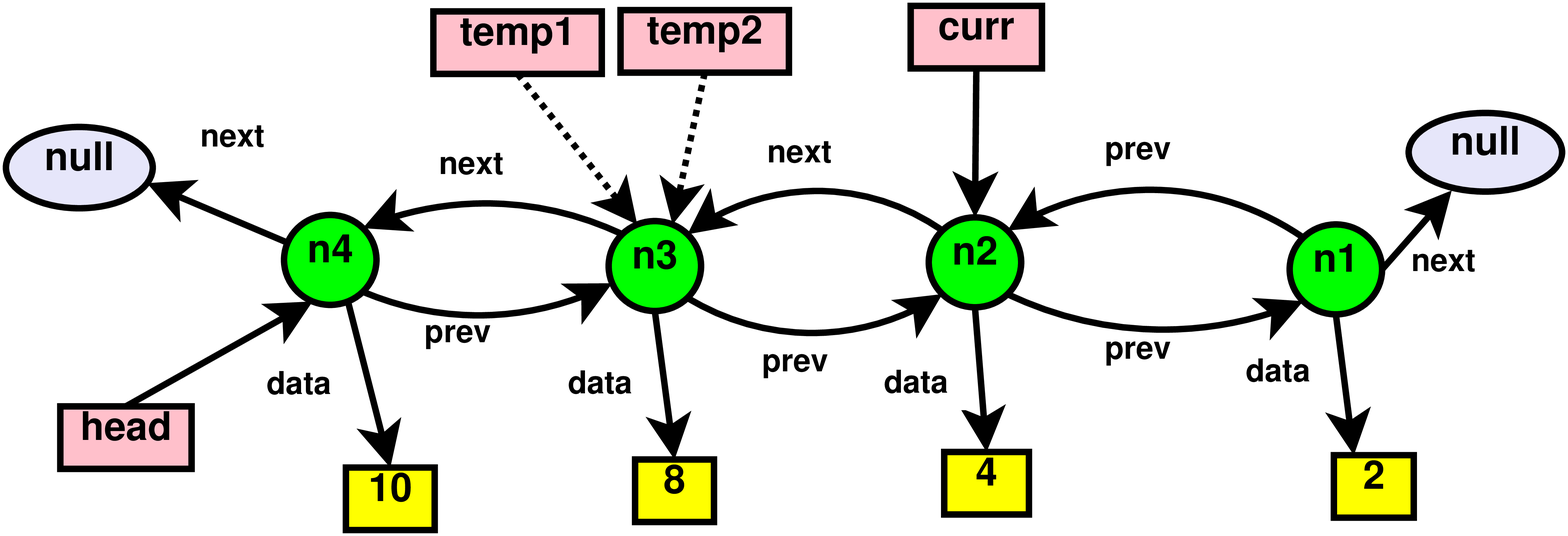}
 \caption{At the end of second loop iteration (after user changes)}
 \label{fig:second}
 \end{subfigure}%
 \\
 \begin{subfigure}[b]{0.49\textwidth}
 \centering
 \includegraphics[width=\textwidth]{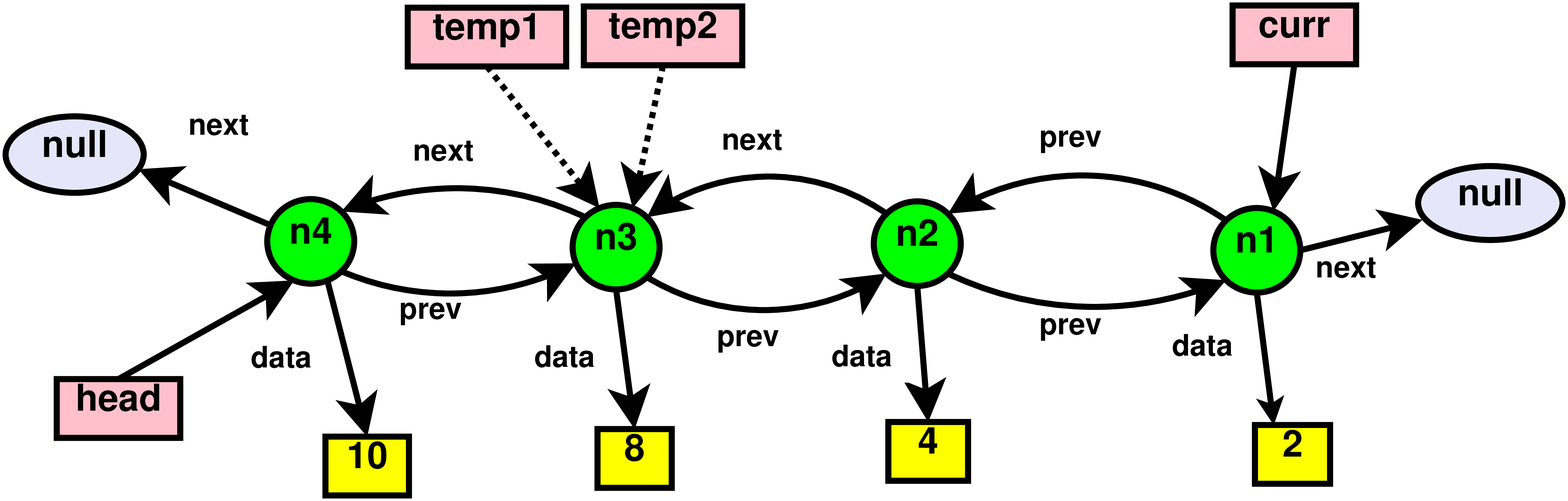}
 \caption{At the end of third loop iteration (no changes required)}
 \label{fig:justbefore}
 \end{subfigure}
 \hfill
 \begin{subfigure}[b]{0.49\textwidth}
 \centering
 \includegraphics[width=\textwidth]{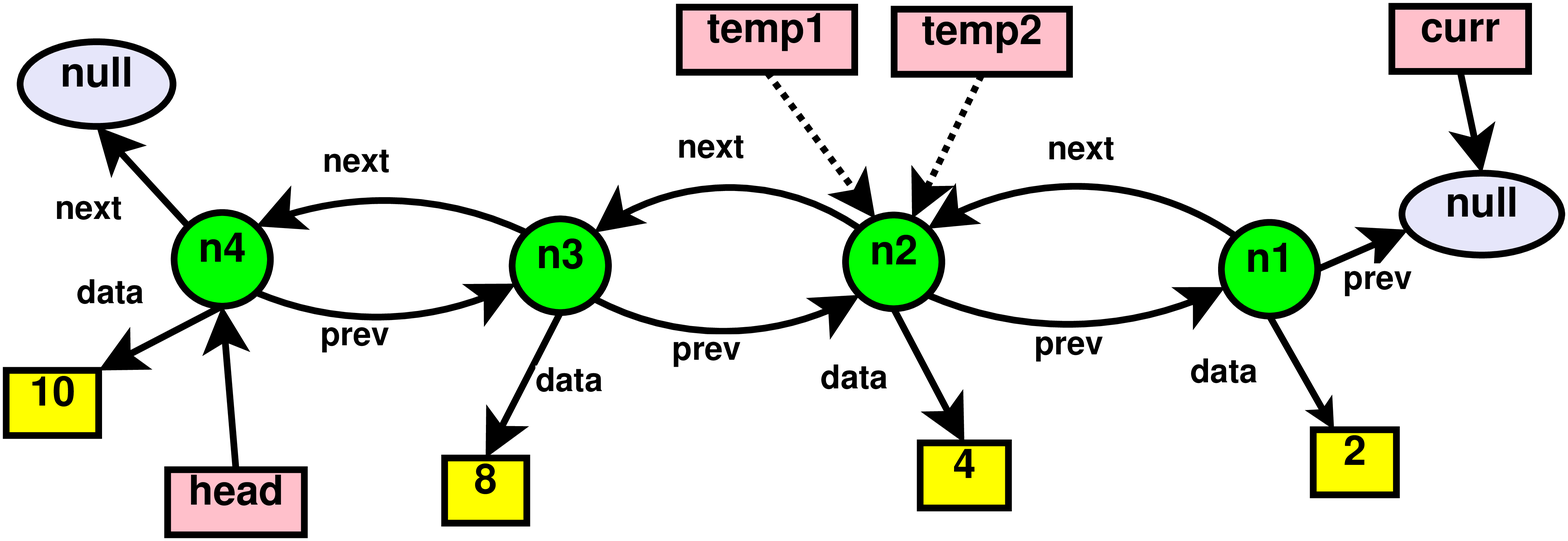}
 \caption{At the end of fourth loop iteration}
 \label{fig:final}
 \end{subfigure}
 \caption{Visualization of the program execution provided by \toolname.}
 \label{fig:full}
 \vspace{-2pt}
\end{figure*}

We demonstrate a typical debug-localize-repair session on \toolname: the program in \fig{fig:motivating} creates a doubly linked-list (stack) of four nodes using the \var{push()} functions, and then, calls the \var{reverse()} function to reverse this list. The \var{reverse()} function contains three faults:
\begin{enumerate}[noitemsep,wide=5pt, leftmargin=25pt]
 \item The loop condition is buggy which causes the loop to be iterated for one less time than expected;
 \item The programmer (possibly due to a cut-and-paste error from the previous line) sets \var{temp2} to the \var{prev} instead of \var{next} field;
 \item The \var{head} pointer has not been set to the new head of the reversed list.
\end{enumerate}

The programmer uses the \underline{\var{start}} command to launch \toolname, followed by four \underline{\var{next}} commands to concretely execute the statements in \var{push()} functions, creating the doubly-linked list. \fig{fig:start} shows the current (symbolic) state of the program heap, that is displayed to the programmer.

\colbox{{\prmt start\\ Starting program...\\ push(2)\\ \prmt next; next; next; next;}}

\colbox{{push(4);\\\dots}}

The programmer, then, uses the \underline{\var{step}} command to step into the \var{reverse()} function.

\colbox{{reverse();\\ \prmt step}\\{current = head; \\ \prmt next}}

%

The programmer deems the currently displayed state as desirable as this program point and decides to assert it via the \underline{\texttt{spec}} command. The asserted states are registered as part of the specification, and the {\it repair module} ensures that any synthesized program repair does exhibit this program state at this program location.


\colbox{{while(temp1 != NULL) \\ \prmt spec\\ Program states added}}

\textit{Bug1} prevents the execution from entering the while-loop body, the programmer therefore employs \underline{\texttt{enter}} command to force the execution inside the loop.

\colbox{{while(temp1 != NULL) \\ \prmt enter}}

The programmer issues multiple \underline{\tt next} commands to reach the end of this loop iteration.

\colbox{temp1 = current->prev;\\ \prmt next; next; next; next; next; \\ \dots\\ while(temp1 != NULL)}

The program state at this point (\fig{fig:first}) seems undesirable as \var{current} and \var{prev} field of node \var{n4} point to \var{null} (instead of pointing to \var{n3}). The programmer \textit{corrects the program state} by bringing about these changes via the \underline{\var{change}} command.


\colbox{\prmt change current n3}\\
\colbox{\prmt change n4 -> prev n3}\\

\fig{fig:change} shows the updated program state, and the programmer commits them to specification. 

\colbox{\prmt spec\\ Program states added}\\

The execution is now forced in the loop for the second time, again using the \underline{\var{enter}} command. 

\colbox{{\prmt enter\\ while(temp1 != NULL)}\\ \dots}\\

The state at the end of the second iteration is not correct; the programmer performs the necessary changes and commits it to the specification.

\colbox{{while(temp1 != NULL) \\ \prmt change current n2 \\ \prmt change n3 -> prev n2} \\ \prmt spec\\ Program states added}\\ 
%
%

She then uses the \underline{\var{repair}} command to request a repair patch.

\colbox{\prmt repair\\ Repair synthesized...}

To repair the program, \toolname first launches its \textit{bug localization module} that searches for potentially faulty statements; in this case, it identifies the second, third, and fifth statements (lines 7, 6, and 10) in the ``while'' loop (which is the statement with \textit{Bug2}) as suspicious candidates.

The repair module, then, searches for possible mutations of the potentially faulty statements (identified by the bug localizer) to synthesize a repair patch that is guaranteed to satisfy the given specifications committed thus far. 

In the present case, the repair synthesized by \toolname correctly fixes \textit{Bug2}; however, the other bugs remain as the trace has not encountered these faults yet. 
\toolname, further, simulates {\it hot-patching} of this repair, allowing the user to continue this debugging session rather than having to abort this debug session, recompile, and restart debugging.

To check the generality of the repair, the programmer steps through the third loop iteration to confirm that it does not require a state change, alluding to the fact that the repair patch is possibly correct.

\colbox{{while(temp1 != NULL)\\ \prmt enter \\ \dots}}\\

The fourth iteration also updates the program heap as per the programmer's expectations, reinforcing her confidence in the repair patch.


Due to \textit{Bug1}, the loop termination condition does not hold even after the complete list has reversed; the programmer, thus, forces a change in the control flow via the \underline{\var{leave}} command to force the loop exit.\\
\colbox{{while(temp1 != NULL)\\ \prmt leave}}\\
\colbox{Exiting function... }


At this point, the programmer notices that the state is faulty as the \var{head} pointer continues to point to the node \var{n4} rather than \var{n1}, the new head of the reversed list (\fig{fig:final}). 

The programmer adds this change to the specification and requests another repair patch.

\colbox{{\prmt change head n1}}\\
\colbox{{\prmt spec\\ Program states added}}\\
\colbox{{\prmt repair \\ Repair synthesized...}}


This repair requires the \textit{insertion} of a new statement; \toolname is capable of synthesizing a bounded number of additional statements to the subject program. On our machine, the first repair call takes 0.5 s (fixing \textit{Bug2}) while the second repair call returns in 0.3 s (fixing \textit{Bug1} and \textit{Bug3}).


To summarize, the debug session builds a correctness specification via \textit{corrections to the program state}, that \toolname uses to drive automated repair, aided by fault localization to prune the repair space. 
\subsection{The Claws of \toolname}

\begin{figure}[t]
        \centering
        \includegraphics[width=\columnwidth]{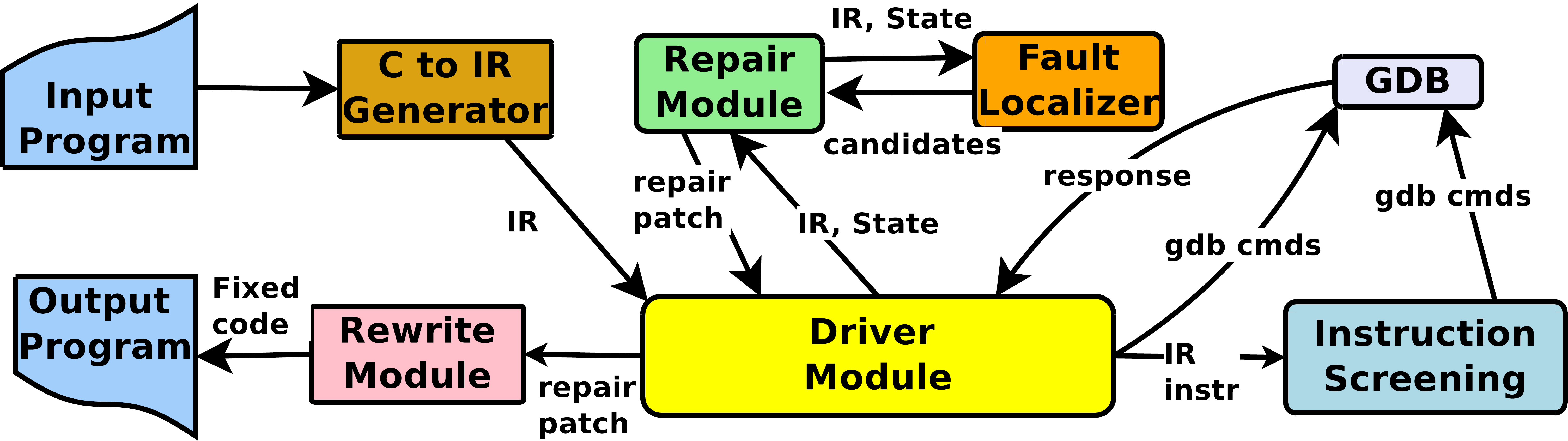}
        \caption{The claws of \toolname}
        \label{fig:eagle}
\end{figure}

The high-level architecture of \toolname is shown in \fig{fig:eagle}. The {\it Driver} module is the heart of the tool, providing the user shell and coordinating between other modules.


After receiving a C program, \toolname employs the {\it C-to-IR} generator to compile it into its intermediate representation (IR) as a sequence of guarded statements ($\Gamma$) and a location map ($\Lambda$) to map each line of the C-source code to an IR instruction (see \sect{sec:algorithms}). 
Each C-source code instruction can potentially be mapped to multiple IR instructions. For the sake of simplicity, we assume that each C-source code line appears in a new line. Note that each C-code instruction can get compiled down to multiple IR instructions.

The {\it Driver} module initiates the debug session by loading the binary on \var{gdb}: many of the commands issued by the programmer are handled by dispatching a sequence of commands to \var{gdb} to accomplish the task. However, any progress of the program's execution (for example, the \underline{{\tt next}} command from the programmer) is routed via the {\it instruction screening} module that manages specification refinement and simulates hot-patching (see \sect{sec:algorithms} and \alg{alg:execute-statement}).

On the \underline{{\tt repair}} command, the {\it driver} invokes the {\it repair module} to request an automated repair based on the specification collected thus far. The repair module, in turn, invokes the \textit{fault localization engine}, to identify a set of suspicious locations. The fault localization algorithm is sound but not complete---though it may return multiple suspicious statements (including ones that are not faulty), the set of these suspicious locations is guaranteed to contain the buggy location. The repair module restricts its mutations within the set of suspicious statements to synthesize a repair patch. This patch is propagated to the {\it instruction screening} module to enable hot-patching, enabling the user to continue \textit{as if she was executing this transformed program all along}. If satisfied, she invokes the \textit{rewrite} module to translate the intermediate representation of the repaired program to a C language program.

\section{Heap Debugging}
\label{sec:algorithms}

The state of a program ($\mathcal{S}$) contains a set of variables $\sigma_{V}$ and a set of heap nodes $\sigma_{H}$ with fields $\sigma_{F}$ as $\mathcal{V} \times \mathcal{H}$; the state of the program variables, $\mathcal{V}$, is a map $\sigma_{V} \rightarrow \mathcal{D}$ and the program heap ($\mathcal{H}$) is a map $\sigma_{H} \times \sigma_{F} \rightarrow \mathcal{D}$. 
The domain of possible values, $\mathcal{D}$, is $\mathcal{I} \cup \sigma_{H}$ where $\mathcal{I}$ is the set of integers. 
For simplicity, we constrain the discussions in this paper to only two data-types: integers and pointers. We use the function $\Upsilon(e)$ to fetch the type of a program entity; a program entity $e \in \mathcal{E}$ is either a variable $v \in \sigma_{V}$ or a field of a heap node $h\in \sigma_{H}\times \sigma_{F}$. Also, pointers can only point to heap nodes as we do not allow taking reference to variables. 

Memory state witnessed by concrete execution via \var{gdb} is referred to as the concrete state, from which we extract the symbolic state as a memory graph~\cite{Zimmermann:2001}, where machine addresses are assigned symbolic names. For our \textit{symbolic} state, pointers are maintained in symbolic form, whereas scalar values (like integers) are maintained in concrete form. In the \textit{concrete} state, all entities are maintained in their concrete states.


\subsection{Symbolic Encoding of an Execution}

\begin{figure*}[t]
        \centering
          \begin{subfigure}[t]{0.42\textwidth}
          \begin{lstlisting}[language = C]
if (head != NULL) {
  temp = head->next;
}
if (head != NULL 
     || temp != NULL) {
  return;
}
last = head;
temp = last->next; \end{lstlisting}
        \caption{An example program in C}
        \end{subfigure}%
          ~ 
        \begin{subfigure}[t]{0.58\textwidth}
          \centering
          \begin{lstlisting}[language = C]
1 (true) : b0 = (not (head == null))
2 (b0)   : temp = head.next
3 (true) : b1 = (head == null)
4 (true) : b2 = (temp == null)
5 (true) : b1 = b1 bor b2
6 (b1)   : halt
7 (true) : last = head
8 (true) : temp = last.next \end{lstlisting}
      \caption{Intermediate representation of the program}
      \end{subfigure}
\caption{Our Intermediate Representation}
\label{fig:example_IR}
\end{figure*}

We use Hoare triples~\cite{Hoare:1969} to show the semantics of our intermediate representation (\fig{fig:fwd_semantics}). In the intemediate representation, the program is maintained as a sequence of guarded statement, i.e. a statement is executed only if its guard evaluates to {\tt true} at that program point (rules {\tt grd1} and {\tt grd2} in \fig{fig:fwd_semantics}). The scope of our repairs includes modification/insertion of both statements and guards.


Assignment~\mbox{($x:=y$)}, getfield~\mbox{($x:=y.f$)} and putfield~\mbox{($x.f = y$)} are the primary statements in our intermediate representation. For a map $\mathcal{M}$, we use the notation $\mathcal{M}_1=\mathcal{M}_2[a_1\mapsto a_2]$ to denote that $\mathcal{M}_2$ inherits all mappings from $\mathcal{M}_1$ except that the mappings $a_1\mapsto a_2$ is added/updated. For brevity, we skip discussions of other statements like \var{print}.

\begin{figure*}[t]
  \scalebox{0.8}{
 \begin{minipage}{\textwidth}
 \hspace{3em}\infer[asgn]
 {\opsemf{\langle \mathcal{V}_1, \mathcal{H}_1\rangle}{x\asgn y}{\langle \mathcal{V}_2, \mathcal{H}_1\rangle}}
 {\Upsilon(x)=\Upsilon(y) & \mathcal{V}_2=\mathcal{V}_1[x\mapsto \mathcal{V}_1(y)]}
 
 \begin{align*}
 \infer[getfld]
 {\opsemf{\langle \mathcal{V}_1, \mathcal{H}_1\rangle}{x\asgn y.f}{\langle \mathcal{V}_2, \mathcal{H}_1\rangle}}
 {\Upsilon(x)=\Upsilon(y.f) \quad \Upsilon(y)=\text{ptr} \quad \mathcal{V}_1(y)\neq\text{null} \\ \hfill \mathcal{V}_2=\mathcal{V}_1[x\mapsto \mathcal{H}_1\langle\mathcal{V}_1(y),f\rangle]\hfill} \hspace{1.5em}
 \infer[putfld]
 {\opsemf{\langle \mathcal{V}_1, \mathcal{H}_1\rangle}{x.f\asgn y}{\langle \mathcal{V}_1, \mathcal{H}_2\rangle}}
 {\Upsilon(x)=\Upsilon(y.f) \quad \Upsilon(x)=\text{ptr} \quad \mathcal{V}_1(x)\neq \text{null}\\ \hfill \mathcal{H}_2=\mathcal{H}_1[\langle\mathcal{V}_1(x),f\rangle \mapsto \mathcal{V}_1(y)]\hfill \ }
 \end{align*}%
 
 \begin{align*}
 \infer[cncr]
 {\opsemf{\langle \mathcal{V}_1, \mathcal{H}_1\rangle}{concrete(\zeta)}{\langle \mathcal{V}_2, \mathcal{H}_1\rangle}}
 {\hfill \hoarerule{\langle \overline{\mathcal{V}}_1, \overline{\mathcal{H}}_1\rangle}{\zeta}{\langle \overline{\mathcal{V}}_2, \overline{\mathcal{H}}_2 \rangle} \hfill\\ \quad \mathcal{V}_2, \mathcal{H}_2 = Sym(\overline{\mathcal{V}}_2, \overline{\mathcal{H}}_2) \quad \overline{\mathcal{V}}_1, \overline{\mathcal{H}}_1 = Concr(\mathcal{V}_1, \mathcal{H}_1)} \hspace{1.5em} 
 \infer[skip]
 {\opsemf{\langle \mathcal{V}_1, \mathcal{H}_1\rangle}{skip}{\langle \mathcal{V}_1, \mathcal{H}_1\rangle}}
 {}
 \end{align*}%
 
 \begin{align*}
 \infer[grd1]
 {\opsemf{\langle \mathcal{V}_1, \mathcal{H}_1\rangle}{grd\text{ ? }stmt}{\langle \mathcal{V}_2, \mathcal{H}_2\rangle}}
 {[\![ grd ]\!]=\text{true} & \hoarerule{\langle \mathcal{V}_1, \mathcal{H}_1\rangle}{\text{stmt}}{\langle \mathcal{V}_2, \mathcal{H}_2\rangle}} \hspace{1.5em}
 \infer[grd2]
 {\opsemf{\langle \mathcal{V}_1, \mathcal{H}_1\rangle}{grd\text{ ? }stmt}{\langle \mathcal{V}_1, \mathcal{H}_1\rangle}}
 {[\![ grd ]\!]=\text{false}}
 \end{align*}%
 
 \begin{align*}
 \infer[nondet]
 {\opsemf{\langle \mathcal{V}_1, \mathcal{H}_1 \rangle}{x\asgn nondet}{\langle \mathcal{V}_2, \mathcal{H}_1 \rangle}}
 {\mathcal{V}_2=\{y\mapsto h\ | \ (h \overset{*}{\gets} \mathcal{H}_1, y=x) \land ((y\mapsto h) \in \mathcal{V}_1, y\neq x)\}} \hspace{1.5em}
 \infer[band]
 {\opsemf{\langle \mathcal{V}_1, \mathcal{H}_1\rangle}{z \asgn x \hspace{3pt} band \hspace{3pt} y}{\langle \mathcal{V}_2, \mathcal{H}_2\rangle}}
 {{\Upsilon(x)=\Upsilon(y)} & \\ \mathcal{V}_2=\mathcal{V}_1[z\mapsto \mathcal{V}_1(x) \wedge \mathcal{V}_1(y)]}
 \end{align*}%
 
 \begin{align*}
 \infer[bor]
 {\opsemf{\langle \mathcal{V}_1, \mathcal{H}_1\rangle}{z \asgn x \hspace{3pt} bor \hspace{3pt} y}{\langle \mathcal{V}_2, \mathcal{H}_2\rangle}}
 {\Upsilon(x)=\Upsilon(y) & \mathcal{V}_2=\mathcal{V}_1[z\mapsto \mathcal{V}_1(x) \vee \mathcal{V}_1(y)]} \hspace{1.5em}
 \infer[bassign]
 {\opsemf{\langle \mathcal{V}_1, \mathcal{H}_1\rangle}{z = x \hspace{3pt} \star \hspace{3pt} y}{\langle \mathcal{V}_2, \mathcal{H}_2\rangle}}
 {\Upsilon(x)=\Upsilon(y) & \mathcal{V}_2=\mathcal{V}_1[z\mapsto [\mathcal{V}_1(x) \star \mathcal{V}_1(y)]]\\ &\star \in \{=, \neq, <, >, \leq, \geq\}}
 \end{align*}%
 
 \end{minipage}
  }
 \caption{\textbf{Forward Semantics} of our intermediate representation}
 \label{fig:fwd_semantics}
\end{figure*}
 
Each statement comes with its set of preconditions that must be satisfied for the update to ensue, for example, the assignment statement requires that the type of the variables match, and a getfield statement requires type matching along with the need for the dereferenced variable to be a pointer and non-\var{null}. 


We use \var{concrete} statement to allow an interesting debugging strategy (referred to as \textit{specification refinement}, see \sect{sec:spec_refine}). For such statements, $\zeta$, we extract a concrete precondition, execute $\zeta$ \textit{concretely} and fetch a concrete postcondition for \var{concrete} statement.


The guards are predicates that can involve comparisons from \{$\leq, <, \geq, >, =, \neq$\} for integers and only \{$=, \neq$\} for pointers (we omit their formal semantics).

The boolean predicates for the guards can be constructed using the boolean assignment statement ($z = x \star y$, where $\star \in \{=, \neq, <, >, \leq, \geq\}$); compound conditions over these predicates are constructed via the boolean-or operator ($z := x\ \textrm{bor}\ y$) or the boolean-and operator ($z:=x\ \textrm{band}\ y$). The program also allows for \textit{non-deterministic} assignments ($x:= nondet$). 
\fig{fig:example_IR} shows a program with compound conditions and its intermediate representation. 

When the \underline{{\tt repair}} command is invoked at a particular program point $p$, by the conjunction of the semantic encoding of the instructions in the trace, the \textit{repair module} constructs a symbolic model, $\Phi_{sem}$, of the execution trace upto~$p$ (we assume that the trace has a length of \textit{n}).


{\centering
 $ \displaystyle
 \begin{aligned} 
\Phi_{sem} \equiv \prod^n_{i=1} \mathcal{T}_i(\mathcal{S}_i, \mathcal{S}_{i+1})
\end{aligned}
 $ 
\par}

\noindent $\mathcal{T}_i$ encodes the semantics of the $i^{th}$ instruction (\fig{fig:fwd_semantics}) and $\mathcal{S}_i$ (and $\mathcal{S}_{i+1}$) denote the input (and output) state of this instruction (respectively).


\subsection{The Heap Debugger}


\RestyleAlgo{boxruled}
\begin{algorithm}[t]
\caption{\label{alg:heap-debugger}The Heap Debugger}
$\Theta = \{e | e \in \mathcal{H}\}$\\
\While{true}{
 $cmd \asgn Prompt()$\\
 \Switch{cmd}{
 \uCase{\texttt{start}}{loc = gdb\_start()}
 \uCase{\texttt{next}}{loc = ExecuteStatement(\texttt{loc})\\ $\st_c$ = fetch\_concrete\_state() \\ $\st_s, \gamma$ = create\_symbolic\_state($\st_c, \gamma$) \\ display\_map($\st_s$)}
 \uCase{\texttt{break <loc>}}{gdb\_send(\texttt{break <loc>)}}
 \uCase{\texttt{change <$e_s$> <$v_s$>}}{$\st_c$ = fetch\_concrete\_state() \\ $\st_s, \gamma$ = create\_symbolic\_state($\st_c, \gamma$) \\ $\st_s[e_s]\asgn v_s$ \\ $gdb\_set\_address(\gamma[e_s], v_s)$}
 \uCase{\texttt{spec}}{$\st_c$ = fetch\_concrete\_state() \\ $\st_s, \gamma$ = create\_symbolic\_state($\st_c, \gamma$) \\ $assert\_spec(\Lambda[loc].IR\_id, \st_s \cap \Theta)$}
 \uCase{\texttt{repair}}{$\var{patch} \asgn repair\_run()$\\ hot\_patch(\var{patch})}
 }
}
\end{algorithm}

\RestyleAlgo{boxruled}
\begin{algorithm}[t]
 \KwIn {$\Gamma :: [\langle stmt.action, stmt.grd, stmt.loc]$, \\ \hspace{30pt}$\Lambda :: N \rightarrow \langle L, \{changed, inserted, preserved\} \rangle$} 
 \caption{\label{alg:execute-statement}ExecuteStatement}
 \uIf{stmt == "concrete"}{
 loc = gdb\_send("next")\\
 $\st_c$ = fetch\_concrete\_state() \\ $\st_s, \gamma$ = create\_symbolic\_state($\st_c, \gamma$) \\
 repair\_add\_spec($\Gamma[pp].IR\_id, \st_s \cap \Theta$)
 } \uElseIf{$\Lambda[loc].status == changed$} {
 gdb\_send("skip")\\
 irstm = IR2gdbStm($\Lambda[loc].IR\_id$)\\
 gdb\_exec\_stm(irstm)
 } \uElseIf{$\Lambda[loc].status == inserted$} {
 irstm = IR2gdbStm($\Lambda[loc].IR\_id$)\\
 gdb\_exec\_stm(irstm)
 } \lElse {
 loc = gdb\_send("next")
 }
 \Return loc
\end{algorithm}

\alg{alg:heap-debugger} provides a high-level sketch of the operation of \toolname. Our algorithm accepts a program as a sequence of guarded statements $\langle \pi, \omega \rangle$ where $\pi$ is a {\it guard} predicate of the form $\langle op, arg1, arg2 \rangle$, and $\omega$ can be one of assignment, getfield, putfield, boolean-or, boolean-and, booelan-assign, non-deterministic or concrete statement. 

Before the algorithm enters the command loop, $\Theta$ is initialized to the set of entities in the program; these entities are recorded whenever \underline{\var{spec}} is invoked.


When \underline{\var{next}} command is invoked, \toolname dispatches the next program statement to be executed (at source line \var{loc}) to the {\it statement screening module} (\alg{alg:execute-statement}), which returns the next C-code source line to be executed. After execution, \toolname uses the function \var{fetch\_concrete\_state()} to query {\tt gdb} for the updated program state, and then generates the memory map~\cite{Zimmermann:2001} (using the \var{create\_symbolic\_state()} function). 
This function returns the symbolic map $\mathcal{\widehat{S}}$ and a map $\gamma$; the map $\gamma$ records the mapping of the concrete entities to their symbolic forms. The symbolic memory map is displayed to the programmer as a visual aid for debugging (see \sect{sec:introduction}).

The functionality of \underline{\var{break}} command is similar to the default instruction in {\tt gdb}: for inserting a breakpoint.


The \underline{\var{change}} command allows the programmer to convey expectations of the desired program state at a point. The program state can be changed by providing a new value $v_s$ to a symbolic state $e_s$. \toolname fetches the relevant concrete state and issues multiple commands (summarized by the function \texttt{gdb\_set\_address()}) to modify the concrete program state.


The \underline{\var{spec}} command asserts the symbolic state at the current program point. 

Finally, the \underline{\var{repair}} command invokes the {\it repair module}, which attempts to synthesize a repair patch that satisfies the concatenated assertions added thus far. The {\it repair module} internally calls the fault localizer and, hence, fault localization is not available as an explicit command. If the repair succeeds, the repair patch is passed to the {\it instruction screening} module, which simulates hot-patching for a seamless debugging session.

In addition to the above, \toolname also supports altering of the control flow (like entry/exit of loops via the \underline{\var{enter}} and \underline{\var{leave}} commands respectively), flip branch directions etc. We demonstrated these features in \sect{sec:introduction} but we omit the details.

The \texttt{ExecuteStatement()} function, provided by the \textit{instruction screening} module, is described in \alg{alg:execute-statement}; this function accepts a list of guarded statements $\Gamma$ and a map $\Lambda$ from the source line numbers (in~$N$) to a tuple containing the corresponding IR instruction(s) (in~$L$). 
Status bits ($F\in \{changed, inserted, preserved\}$) are used to indicate if: (a) the IR instruction has been modified ($changed$) by a repair patch, (b) appears as a new instruction ($inserted$) due to a repair patch, or (c) is unmodified ($preserved$); this information is required to simulate hot-patching. 
Deletion of a statement is also marked ($changed$); the corresponding patch simply sets the guard of the instruction to \var{false}. 
The \textit{instruction screening} module handles two primary tasks:

\begin{itemize}[noitemsep,wide=5pt, leftmargin=25pt]
 \item {\bf Handling \var{concrete} statements.} Upon encountering a \var{concrete} statement, \toolname executes it via \var{gdb} by issuing the \underline{\var{next}} command. The effect of this concrete execution is then asserted by taking a snapshot of the updated concrete state (again via {\tt gdb}) and adding the corresponding symbolic state to the specification. We refer to this debugging strategy as {\it specification refinement} (see \sect{sec:spec_refine}).
 \item {\bf Simulate hot-patching.} If a repair patch has modified the statement, \toolname requests \var{gdb} to skip the execution of this statement, and translates the ``effect" of the modification into a string of \var{gdb} commands ($irstm$) via the \var{IR2gdbStm()} function and consigns the command-list to \var{gdb} using the \var{gdb\_exec\_stm()} function. 
\end{itemize}

Otherwise, the next statement is concretely executed via \var{gdb} by issuing the \underline{\var{next}} command.

\section{Proof-Guided Repair}
\label{sec:Repair}

\begin{algorithm}[t]
 \caption{\label{alg:repair}Unsat Core Guided Repair Algorithm}
 $\Phi_{grd} = \Phi_{stm} = \Phi_{ins} = \emptyset$\\
 $n \asgn |\porg| + num\_insert\_slots$\\
 \tcc{Assert the input (buggy) program}
 \For{$i\in \{1 \dots |\porg|\}$}{
 \eIf{$i\in \mathcal{L}$}{
 $\Phi_{grd} \pluseq \langle \prep.grd[\xi(i)] == \porg.grd[i] \rangle$\\
 $\Phi_{stm} \pluseq \langle \prep.stm[\xi(i)] == \mathcal{P}.stm[i] \rangle$\\
 }{
 $\Phi_{grd} \pluseq \langle \lnot r_{\xi(i)} \implies (\prep.grd[\xi(i)] == \porg.grd[i]) \rangle$\\
 $\Phi_{stm} \pluseq \langle \lnot s_{\xi(i)} \implies (\prep.stm[\xi(i)] == \mathcal{P}.stm[i]) \rangle$\\
 }
 }
 \tcc{Initialize the insertion slots}
 \For{$i\in \{|\porg| \dots n\}$}{
 $\Phi_{ins} \pluseq \langle \lnot t_{\xi(i)} \implies (\phi_{\xi(i)} == \var{false}) \rangle$\\
 }
 \tcc{Define the placing function $\xi$}
 $\Phi_{bk} \asgn \forall_{i \in \{1 \dots n\}} (1 \leq \xi(i) \leq n) \land \var{distinct}({\xi(i)}) $\\
 $\Phi_{bk} \pluseq \forall_{\langle i,.,. \rangle , \langle k,.,. \rangle \in \mathcal{P}} (i < k \implies \xi(i) < \xi(k))$\\
 $v \asgn \textsc{UNSAT}$\\
 \tcc{Relax till specification is satisfied}
 $\tau_{grd} \asgn \tau_{stm} \asgn \tau_{ins} \asgn 0$\\
 \While{$v = \textsc{UNSAT}$ {\bf or} tries exceeded}{
 $\langle res, \prep, uc \rangle \asgn \textsc{Solve}(\Phi_{spec} \land \Phi_{sem} \land \Phi_{bk},$ \\
 \hspace{60pt} $\Phi_{grd} \land \Sigma_{k \in \{1\dots |\porg|\}} r_k < \tau_{grd},$ \\ 
 \hspace{60pt} $ \Phi_{stm} \land \Sigma_{k \in \{1\dots |\porg|\}} s_k < \tau_{stm}$ \\ 
 \hspace{60pt} $ \Phi_{ins} \land \Sigma_{k \in \{|\porg|+1 \dots n\}} t_k < \tau_{ins})$\\
 \tcc{Use the UNSAT core to drive relaxation}
 \uIf{$res = \textsc{UNSAT}$}{
 \lIf{$\Phi_{grd} \cap uc \neq \emptyset$}{
 $\tau_{grd} \pluseq 1$
 }
 \lElseIf{$\Phi_{stm} \cap uc \neq \emptyset$}{
 $\tau_{stm} \pluseq 1$
 }
 \lElseIf{$\Phi_{ins} \cap uc \neq \emptyset$}{
 $\tau_{ins} \pluseq 1$
 }
 \lElse{
 \Return{null}
 }
 }
 }
 \lIf{tries exceeded}{
 \Return{null}
 }
 \Return{$\prep$}
\end{algorithm}

\alg{alg:repair} shows our repair algorithm: it takes a (buggy) program $\porg$ as a sequence of guarded statements, a set of \textit{locked locations} $\mathcal{L}$, and a bound on the number of new statements that a repair is allowed to insert (\var{num\_insert\_slots}). The repair algorithm attempts to search for a repair candidate $\prep$ (of size $n = |\mathcal{P}| + \var{num\_insert\_slots}$) that is ``close" to the existing program and satisfies the programmer's expectations (specification). Our algorithm is allowed to mutate and delete existing statements and insert at most $n$ new statements; however, mutations are not allowed for the locations contained in $\mathcal{L}$. The insertion slots contain a guard \var{false} to begin with~(Line 8); the repair algorithm is allowed to change it to ``activate" the statement. Deletion of a statement changes the guard of the statement to \var{false}.

\toolname allows for new nodes and temporary variables by providing a bounded number of additional (hidden) nodes/temporaries, made available on demand. The programmer configures the number of insertion slots, but these slots are activated by the repair algorithm only if needed. For loops, we add additional constraints so that all loop iterations encounter the same instructions.

\subsection{Primary Constraints}
We use a set of selector variables $\{r_1, \dots, r_n, s_1, \dots, s_n\}$ to enable a repair. Setting a selector variable to \var{true} relaxes the respective statement, allowing \toolname to synthesize a new guard/statement at that program point to satisfy the specification. We define a metric, $closeness(\mathcal{P}_1, \mathcal{P}_2)$, to quantify the distance between two programs
by summing up the set of guards and statements that match at the respective lines. As the insertion slots should be allowed to be inserted at any point in the program, the closeness metric would have to be `adjusted' to incorporate this aberration due to insertions. For this purpose, our repair algorithm also infers a relation $\xi$ that maps the instruction labels in the repair candidate $\prep$ to the instruction labels in the original program $\porg$; the instruction slots are assigned labels from the set $\{|\porg| +1, \dots, n\}$. We define our closeness metric as:

\vspace{2pt}

\begin{align*}
    \var{closeness}(\porg, \prep) = & \sum_{i=1}^{|\porg|} (\porg.grd[i] = \prep.grd[\xi(i)])\\ & + \sum_{i=1}^{|\porg|} (\porg.stm[i] = \prep.stm[\xi(i)])\\ & + \sum_{i=|\porg| + 1}^{n}(\prep.grd[\xi(i)] \neq \var{false})
\end{align*}
\vspace{2pt}

The above metric weights a repair candidate by the changes in the statements/guards and new statements added (insertion slots {\it activated}).

\alg{alg:repair} starts by asserting the input program $\porg$, via the selector variables if it is not a locked location, as part of the constraints $\Phi_{grd}$ and $\Phi_{stm}$ (lines~3--11), and initializes the insertion slots to their deactivated state (lines~12--14) with selector variables $t_i$. The constraint $\Phi_{bk}$ ensures that the function $\xi$ is well-formed: for each $i$, $\xi(i)$ is a distinct value in the range $\{1\dots n\}$ and is a monotonically increasing function (this ensures that the statements preserve the same order in $\prep$ as the order in $\porg$).

Finally, it uses issues a \textsc{Solve()} query to an SMT solver to solve the repair constraint; the sub-constraint $\Phi_{sem}$ contains the semantic encoding of our intermediate statements (\fig{fig:fwd_semantics}) and $\Phi_{spec}$ contains the specification collected during the debugging session as a result of the \underline{\var{spec}} commands.

\subsection{Proof-Guided Search Space Widening}

To ensure that the repaired candidate program $\prep$ is \textit{close} to the original program, we progressively relax the closeness bounds. The variables $\tau_{grd}, \tau_{stm},$ and $\tau_{ins}$ constrain the distance (in terms of changed guards, statements and activated insertion slots) of a repair candidate from the original program.

We use a {\it Proof-Guided Repair Strategy}: the unsat core ($uc$) produced from the proof of unsatisfiability directs us to the bound that needs to be relaxed. The unsat core represents the central {\it reason} as to why the program cannot be made to satisfy the specification; if a constraint $\langle s_i \implies \dots \rangle$ is found in the unsat core, it implies that the reason for unsatisfiability {\it may} be attributed to the fact that $s_i$ is false! Hence, one possible way to remove this unsatisfiability is to increase the bound on $\tau_{stm}$ that allows $s_i$ to turn \var{false}. 

At the same time, we would also like to enforce a priority on the relaxations; for instance, deletion of a statement or mutation of a guard can be considered ``smaller" changes than changing a statement, or worse, inserting a new statement. The chain of conditions (lines~25--27) ensures that, if the unsat core directs us to a possibility of smaller change, we relax the respective bound before others. Finally, on a successful repair, we return the repaired program~$\prep$.

\noindent Guiding repair via the unsat proof has multiple advantages:
\begin{itemize}[noitemsep,wide=5pt, leftmargin=15pt]
 \item The unsatisfiability core (uc) guides us to a feasible repair; for example, if $uc$ does not contain the constraints pertaining to activation of the insertion slots, then it is unlikely that inserting a new statement will fix the bug;
 \item It allows us to prioritize the repair actions; one would prefer mutation of a statement than the insertion of a new statement;
 \item The strategy is fast as the solver operates on a constrained search space that is incrementally widened (in a direction dictated by the proofs) as the search progresses. In case the program to be repaired is close to the original program, the solver will be provided only ``easy" instances that are allowed to mutate/insert a small number of statements;
 \item It allows a fail-fast (line 28) if the specification is buggy or the repair is not possible due to structural constraints (like the number of insertion slots provided); if $uc$ does not contain any constraint from \{$\Phi_{grd}, \Phi_{stm}, \Phi_{ins}$\}, then the program cannot be repaired via any repair action without violating the hard constraints (like the program semantics).
\end{itemize}

The unsat core not only identifies the possible culprits (a sort of bug localization) but also allows us to define a priority among our repair preferences. 
To the best of our knowledge, ours is the first repair algorithm that uses unsat proofs to direct repair; however, this idea has threads of similarity with a model-checking algorithm, referred to as underapproximation widening\cite{Grumberg:2005} (see \sect{sec:related}). 

We evaluated a variant (\textbf{AlgVar}) of our proof directed repair scheme: instead of increasing the respective repair bound, we randomly relax one of the constraints from the unsat core. However, we found that the unsat cores are poor---quite far from the minimum unsat core. Hence, this variant of our algorithm performs poorly, both in terms of success rate and the time taken for repair (see \sect{sec:experiments}).




\section{Bug localization}
\label{sec:localization}
The objective of our bug localization module is to identify a (small) set of statements that are likely to contain the fault(s). 
Our algorithm is targeted at localizing faults for use by the repair phase of \toolname: our localization algorithm localizes faults on concrete program traces using the assertions as precondition/postcondition pairs. 

The bug localization phase exposes two primitives to the repair phase:

\begin{itemize}
 \item Statement locks: Adding a ``locked" attribute to a statement asserts the statement in its position;
 \item Non-deterministic assignment: A non-deterministic assignment allows us to assign an angelic value.
\end{itemize}

\begin{definition}
    {\bf (Upward exposed statement)} A statement whose left-hand side expression (variable or field definition) or its alias has not been assigned by any preceding program statement.
\end{definition}
\begin{definition}
    {\bf (Downward exposed statement)} A statement whose left-hand side expression (variable or field definition) or its alias has not been assigned by any following program statement.
\end{definition}

\begin{definition}
    {\bf (Sandwiched statement)} A statement that is neither upward exposed nor downward exposed.
\end{definition}



\subsection{Intuition}
In this section, we provide the intuition behind our localization algorithm with a few examples.

\subsubsection{Program with a single bug and semantically independent statements}


\begin{table}[t]
        \centering
        \caption{Single bug in a program with semantically independent statements. 
        We show states as a tuple with the values of variables $\B{a,b,c,d}$.} 
        \begin{tabularx}{0.95\textwidth}{ C | C || c | c || C }
        \toprule
       \rowcolor{lightgray}
            Forward & Backward & Diff. & Grad. & Correct Execution \\
        \midrule
        \midrule
            \makecell{\textcolor{blue}{(precondition)}\\ $\B{n_1, n_2, n_3, n_5}$} & $\B{n_1, n_2, n_3, n_5}$ & 0 &  &  \\ \hline
       \rowcolor{arylideyellow}
             \multicolumn{3}{c}{a = a.next} & 0 & \\ \hline
            $\B{n_2, n_2, n_3, n_5}$ & $\B{n_2, n_2, n_3, n_5}$ & 0 & & $\B{n_2, n_2, n_3, n_5}$ \\ \hline
       \rowcolor{arylideyellow}
             \multicolumn{3}{c}{b = b.next \textcolor{blue}{(fix: b = d)}} & 1 & \\ \hline
            $\B{n_2, n_3, n_3, n_5}$ & $\B{n_2, n_5, n_3, n_5}$ & \textcolor{red}{1} & & $\B{n_2, n_5, n_3, n_5}$  \\ \hline
       \rowcolor{arylideyellow}
             \multicolumn{3}{c}{c = c.next} & 0 & \\ \hline
            $\B{n_2, n_3, n_4, n_5}$ & \makecell{\textcolor{blue}{(postcondition)}\\ $\B{n_2, n_5, n_4, n_5}$} & \textcolor{red}{1} & & $\B{n_2, n_5, n_4, n_5}$  \\ 
        \bottomrule
        \end{tabularx}
        \label{tab:single_bug_inde_ue}
\end{table}

Consider the program shown in the second column of \tab{tab:single_bug_inde_ue}: in this case, all the statements are semantically independent, i.e., there does not exist any true dependencies among the statements. To understand the situation, let us also consider the states in the correct execution. 
The reader should be able to understand the state updates for the forward execution intuitively (first column); the details are discussed in \S\ref{fwd_execution}. 

The backward execution (third column) is an attempt at matching the correct execution---we commence from the postcondition, and ``copy" states from the precondition whenever the assigned variable is \textit{upward exposed}. 
Let us illustrate how the backward execution is performed: the backward execution commences from the postcondition and moves towards the precondition (it is essentially the weakest precondition computation, but under the assumption that some of the states could be buggy). 
For arriving at the backward execution state at the third statement, it checks if \var{c} is \textit{upward exposed}, i.e., if it has a preceding assignment statement that assigns to the variable \texttt{c}. 
As there is no such statement, it simply reads the state for \var{c} from the precondition. The same happens for the other statements. In this case, the backward execution exactly matches the states from the execution of the correct program (fourth column).

Now, one can see that the forward execution states agree with backward execution states at statements before the buggy statement, upon whose execution the difference between forward and backward execution builds up. 
The statement where this difference starts to build up is precisely the buggy location.



\begin{table}[t]
        \centering
        \caption{Single bug program with semantically independent but not upward exposed statements. 
        We show states as a tuple with the values of variables $\B{a,b,c,d}$.} 
        \begin{tabularx}{0.95\textwidth}{ C | C || c | c || C }
        \toprule
       \rowcolor{lightgray}
            Forward & Backward & Diff. & Grad. & Correct Execution \\
        \midrule
        \midrule
            \makecell{\textcolor{blue}{(precondition)}\\ $\B{n_1, n_2, n_3, n_5}$} & $\B{n_1, n_2, n_3, n_5}$ & 0 &  &  \\ \hline
       \rowcolor{arylideyellow}
             \multicolumn{3}{c}{b = a.next} & 0 & \\ \hline
            $\B{n_1, n_2, n_3, n_5}$ & $\B{n_1, n_1, n_3, n_5}$, $\B{n_1, n_2, n_3, n_5}$, $\B{n_1, n_3, n_3, n_5}$, $\B{n_1, n_5, n_3, n_5}$ & 0 & & $\B{n_1, n_2, n_3, n_5}$ \\ \hline
       \rowcolor{arylideyellow}
             \multicolumn{3}{c}{b = b.next \textcolor{blue}{(fix: b = d)}} & 1 & \\ \hline
            $\B{n_1, n_3, n_3, n_5}$ & $\B{n_1, n_5, n_3, n_5}$ & \textcolor{red}{1} & & $\B{n_1, n_5, n_3, n_5}$  \\ \hline
       \rowcolor{arylideyellow}
             \multicolumn{3}{c}{c = c.next} & 0 & \\ \hline
            $\B{n_1, n_3, n_4, n_5}$ & \makecell{\textcolor{blue}{(postcondition)}\\ $\B{n_1, n_5, n_4, n_5}$} & \textcolor{red}{1} & & $\B{n_1, n_5, n_4, n_5}$  \\ 
        \bottomrule
        \end{tabularx}
        \label{tab:single_bug_notUE}
\end{table}
       
The procedure differs when a statement is not upward exposed: for the program shown in second column of \tab{tab:single_bug_notUE}, the statements are not \textit{upward exposed} (first and second statement assign to the same variable \var{b}). 
Due to the previous update, it cannot be determined what would be the state when such a statement is run through a backtrackward traversal (as discussed above). 
In such a situation, we split the states, assigning all possible nodes in the data-structure to the concerned variable (here \var{b}), in the hope that at least one of them would be a state agreeing with the execution of the correct program. 

    We can compute the difference between two states by the number of variables whose values \textit{disagree} amongst these states. We can define a gradient of the difference along between two traces $t_1$ and $t_2$ (of same length) by:
   
    \begin{itemize}
        \item Computing a differential trace $\delta t$ by computing the difference between the corresponding states in $t_1$ and $t_2$;
        \item Finding a gradient of the differential trace, $\partial (\delta t)$, by computing $(\partial (\delta t))_i = ||(\delta t)_i - (\delta t)_{i+1}]||$, i.e. computing the distance between two corresponding trace elements. For a trace (sequence) $t$, we use $t_i$ to refer to its $i^{th}$ element.
    \end{itemize}

    For our example, there are points where there is a \textit{non-zero gradient of difference} between states in forward and backward execution at one location--exactly the buggy statement (in the given example).

\subsubsection{Program with a single bug and semantically dependent statements}


\begin{table}[t]
 \centering
 \caption{Single bug program with semantically dependent and not upward exposed statements. We show states as a tuple with the values of variables $\B{a,b,c,d}$.} 
 \begin{tabularx}{0.95\textwidth}{ C | C || c | c || C }
 \toprule
 \rowcolor{lightgray}
     Forward & Backward & Diff. & Grad. & Correct Execution \\
 \midrule
 \midrule
     \makecell{\textcolor{blue}{(precondition)}\\ $\B{n_1, n_2, n_3, n_5}$} & $\B{n_1, n_2, n_3, n_5}$ & 0 &  &  \\ \hline
 \rowcolor{arylideyellow}
      \multicolumn{3}{c}{a = a.next} & 0 & \\ \hline
     $\B{n_2, n_2, n_3, n_5}$ & $\B{n_2, n_2, n_3, n_5}$ & 0  & & $\B{n_2, n_2, n_3, n_5}$ \\ \hline
 \rowcolor{arylideyellow}
      \multicolumn{3}{c}{b = b.next \textcolor{blue}{(fix: b = d)}} & 1 & \\ \hline
     $\B{n_2, n_3, n_3, n_5}$ & $\B{n_2, n_5, n_3, n_5}$ & \textcolor{red}{1} & & $\B{n_2, n_5, n_3, n_5}$  \\ \hline
 \rowcolor{arylideyellow}
      \multicolumn{3}{c}{c = b.next} & 1 & \\ \hline
     $\B{n_2, n_3, n_4, n_5}$ & \makecell{\textcolor{blue}{(postcondition)}\\ $\B{n_2, n_5, n_6, n_5}$} & \textcolor{red}{2} & & $\B{n_2, n_5, n_6, n_5}$  \\ 
 \bottomrule
 \end{tabularx}
 \label{tab:single_bug_depen}
\end{table}

In the program shown in \tab{tab:single_bug_depen}, the third statement is not buggy, but is semantically dependent on the second statement, which is buggy, leads to a non-zero gradient of state difference even at the third statement, as the forward execution's states further diverge from the backward (and correct) execution owing to dependence. 
Hence, both the buggy statement and its dependent statement is added to the set of suspicious statements. 

\subsubsection{Program with multiple bugs and containing semantically dependent statements}
If the bugs occur in the dependent statements, then as discussed above, all of the statements are added to the set of suspicious statements. 
If the bugs are in independent statements, they will create a non-zero gradient independently, and, hence, the faulty statements will be added to the suspicious set.

\subsection{Program transformation applied while invoking \toolname}


\begin{figure*}[t]
        \centering
          \begin{subfigure}[t]{0.55\textwidth}
          \begin{lstlisting}[language = C]
(true) : fast_ptr = temp.next
(fast_ptr != NULL): temp = fast_ptr     
(true) : prev = slow_ptr                
(true) : slow_ptr = slow_ptr.next \end{lstlisting}
        \caption{IR code (loop region)}
        \label{fig:IR_motivating_1}
        \end{subfigure}%
          ~ 
        \begin{subfigure}[t]{0.45\textwidth}
          \centering
          \begin{lstlisting}[language = C]
[(true) : fast_ptr = $\bigstar$]
`\colorbox{arylideyellow}{\tt (fast\_ptr != NULL): temp = fast\_ptr}'
[(true) :  prev = $\bigstar$]
[(true) :  slow_ptr = $\bigstar$] \end{lstlisting}
          \caption{Trandformed IR code after bug localization in~\fig{fig:IR_motivating_1}}
          \label{fig:IR_motivating_2}  
      \end{subfigure}
      \caption{Illustration of program transformation without insert slots}
\end{figure*}

 
\noindent \fig{fig:IR_motivating_2} shows the transformation when localization is performed by \toolname on the intermediate representation (IR), expressed as a sequence of guarded statements. The input program is shown in \fig{fig:IR_motivating_1}. 

The statement identified as suspicious (shown in \colorbox{arylideyellow}{yellow} color) are free to undergo mutations while the remaining statements are \textit{locked} (denoted by enclosing them in square brackets). 
Since these statements do not depend on any buggy statement, they can also be made non-deterministic (shown by $\bigstar$ on the right-hand side of these locked statements). 

However, when the program allows the insertion of statements (the programmer had allowed non-zero insert slots), locked statements cannot be made non-deterministic.
\fig{fig:locked_deterministic_example_2} shows the transformed IR for a program with insert slots (~\fig{fig:locked_deterministic_example_1}) and therefore the locked statements are frozen.


\begin{figure*}[t]
    \centering
    \begin{subfigure}[t]{0.7\textwidth}
    \begin{lstlisting}[language = C]
(true) : fast_ptr = temp.next
(fast_ptr != NULL): temp = fast_ptr.next     
(true) : prev = slow_ptr.next                
(true) : insert_slot \end{lstlisting}
     \caption{IR code (loop region) with insert slot}
     \label{fig:locked_deterministic_example_1}
     \end{subfigure}%
           
        \begin{subfigure}[t]{0.7\textwidth}
      \centering
      \begin{lstlisting}[language = C]
[(true) : fast_ptr = temp.next]
[(fast_ptr != NULL): temp = fast_pt.next]
[(true) : slow_ptr = slow_ptr.next
`\colorbox{arylideyellow}{\tt (true) \ \ : insert\_slot}'  \end{lstlisting}
     \caption{Trandformed IR code after bug localization in ~\fig{fig:locked_deterministic_example_1}}
     \label{fig:locked_deterministic_example_2}
   \end{subfigure}
    \caption{Illustration of program transformation with insert slots}
\end{figure*}


\subsection{Running Example}

\begin{figure}[t]
\begin{minipage}{\columnwidth}
\begin{lstlisting}[language = C]
struct node *head,*slow_ptr,*fast_ptr;
void deleteMid(){
    struct Node *temp,*prev = NULL;
    slow_ptr = head;
    fast_ptr = head;
    temp = fast_ptr->next;
    while (fast_ptr != NULL  &&  temp != NULL){
        fast_ptr = temp->next;
        if (fast_ptr != NULL){                
            temp = fast_ptr; //FIX: fast_ptr->next 
        }
        prev = slow_ptr;        
        slow_ptr = slow_ptr->next;
    }
    temp = slow_ptr->next;
    prev->next = temp;
    delete slow_ptr;
}
...
int main(){
    push(2);  push(4);  push(8);
    deleteMid();
}
\end{lstlisting}
\end{minipage}
\caption{Example code for illustrating bug localization}
\label{fig:motivating_code}
\end{figure}

This section ties all the steps to show how \toolname operates: we illustrate our localization algorithma in a typical repair session. 
The program in \fig{fig:motivating_code} deletes the middle of a singly-linked list; the list is created using a sequence of \var{push()} functions in the \var{main()} function. 

The \var{deleteMid()} function has a bug in the \var{while} loop. The user starts the execution of the program and steps into the \var{deleteMid()} function with the created linked list. 
It has two pointers, \texttt{slow\_ptr} and \texttt{fast\_ptr}, which are both initiated to head of the list. 
In the loop, \texttt{fast\_ptr} moves at a pace double that of \texttt{slow\_ptr} until it points to the last or last but one node. 
The node that \texttt{slow\_ptr} points at this time is deleted from the list.





\begin{figure*}[t]
    \centering
    \includegraphics[width=1.7\textwidth]{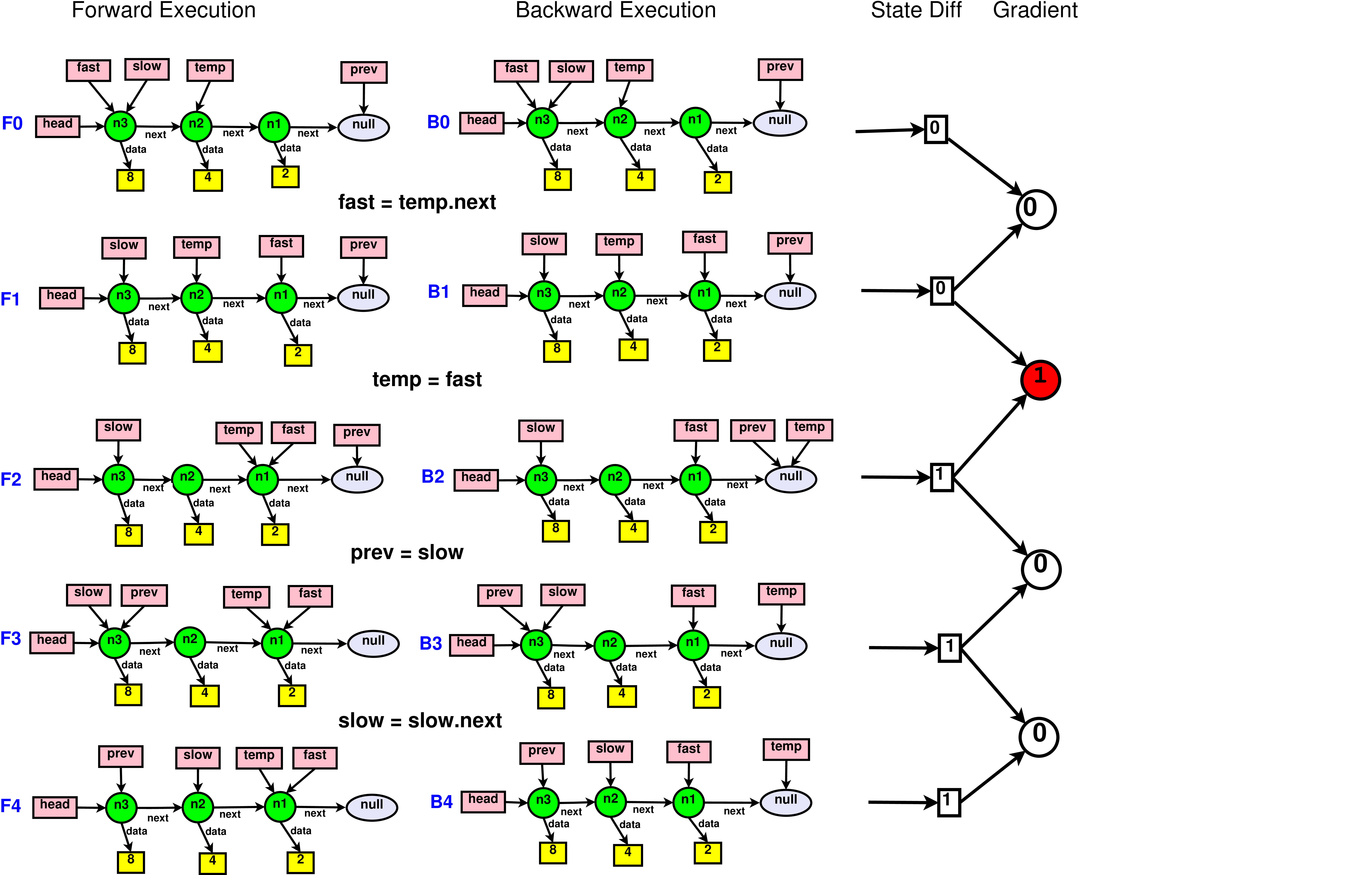}
    \caption{The forward and backward execution states of the program shown in \fig{fig:motivating_code}}
    \label{fig:gradient_local}
\end{figure*}

After the statements before the \var{while} loop are executed, the state displayed to the user is labelled as \textsf{F0} in \fig{fig:gradient_local}. 
The state at this point meets the user's expectation; the user decides to commit it and then enters the loop.
The user next executes all the loop statements. The state at the end of the first iteration is labelled as \textsf{F4} of \fig{fig:gradient_local}. 
Due to the bug in the loop, the obtained state, \textsf{F4}, was not as expected, and hence the user makes a change to the state before the second commit.
The updated state after the change is labelled as \textsf{B4} in \fig{fig:gradient_local}, and user asks \toolname for a repair.

The tool now employs a backward traversal (via the backward semantics) from the correct state, \textsf{B4}, provided by the user, to localize the bug. 
\fig{fig:IR_motivating_1} shows the intermediate representation (IR) on which bug localization module operates. 
The first column in \fig{fig:gradient_local} shows the states in forward execution, and the second column shows the states in backward execution at each statement between the committed states. 
The third column computes the \textit{difference} in the states at respective forward and backward execution (the distance in the map representing the states), while the fourth column shows the \textit{gradient of the difference} of the states corresponding to each program statement.
In this example, only one statement has a non-zero gradient: the second statement of the loop (shown in the red circle in the last column), which is indeed the buggy statement. 

Now, \toolname produces a transformed abstract program (shown in ~\fig{fig:IR_motivating_2}) with a smaller repair space, providing it to the repair algorithm, which synthesizes a repair in a mere 0.2s while the original program required 26s for the repair without localization (speedup of 130$\times$).


\subsection{Algorithm}
We repeat the notations used for the reader's convenience: $\mathcal{S}$ represents a state of a program with a set of variables $\sigma_{V}$ and a set of heap nodes $\sigma_{H}$ with fields $\sigma_{F}$ as $\mathcal{V} \times \mathcal{H}$; the state of the program variables, $\mathcal{V}$, is a map $\sigma_{V} \rightarrow \mathcal{D}$ and the program heap is represented by $\mathcal{H}$ as a map $\sigma_{H} \times \sigma_{F} \rightarrow \mathcal{D}$. 
The domain of possible values, $\mathcal{D}$, is $\mathcal{I} \cup \sigma_{H}$ where $\mathcal{I}$ is the set of integers. 
For simplicity, we constrain the discussions in this paper to only two data-types: integers and pointers. We use the function $\Upsilon(e)$ to fetch the type of a program entity; a program entity $e \in \mathcal{E}$ is either a variable $v \in \sigma_{V}$ or a field of a heap node $h\in \sigma_{H}\times \sigma_{F}$. Also, pointers can only point to heap nodes as we do not allow taking reference to variables. 

\subsubsection{Forward Execution} \label{fwd_execution}

Let $\Delta$ denote the operational semantics in the forward execution.  

$\Delta: State \times statement \to State$

The (repaired) program must satisfy the correctness criterion:
$\Delta(\omega_{pre}, stmt)$ = $\omega_{post}$

where $stmt$ is a program statement and $\omega_{pre}$, $\omega_{post}$ are the precondition (state before executing statement ``stmt") and postcondition (state after executing statement ``stmt"). 

Given a program trace of $n$ statements $[s_0, s_1, \dots, s_n]$, we define $\Delta_i$ as the transition function for the $i^{th}$ statement ($s_i$) in the trace. Subsequently, we denote $\Delta_{i,j}$ to denote the forward transition function for statements from statements $s_i$ to $s_j$.

$\Delta_{i, j}$ = $\Delta_{i} \circ \Delta_{i+1} \circ \Delta_{i+2} \circ \dots \circ \Delta_{j}$

In particular, a transition function for a complete trace of n statements can be written as:

$\Delta_{1, n}$ = $\Delta_{1} \circ \Delta_{2} \circ \Delta_{3} \circ \dots \circ \Delta_{n}$

Forward execution follows the forward semantics (\fig{fig:fwd_semantics}). We illustrate forward execution via \fig{fig:gradient_local}: we use the \var{getfld} rule to execute $fast = temp.next$; the node pointed by \var{fast} is updated to \var{n1} (state \textsf{F1}). We, then, use the \var{asgn} rule to execute $temp = fast$, which changes the node pointed by \var{temp} from \var{n2} to \var{n1}(state \textsf{F2}). 
Then, we again use the \var{asgn} rule to update the value of \var{prev} to \var{n3}, followed by \var{getfld} rule, to revise value of \var{slow} to the node \var{n2} (\textsf{F3} and \textsf{F4} respectively).

\subsubsection{Backward Execution}

We use $\nabla$ to denote the operational semantics for the backward execution.  

$\nabla: State \times statement \to 2^{State}$      

In this case, we may get a set of states instead of a single state (as the same state could be reached by multiple input states).



$\nabla(\omega_{post}, stmt)$ = $\{\omega_{pre_{1}}, \omega_{pre_{2}}, \omega_{pre_{3}}, \dots\}$


$\nabla_{n, 0}(\omega_{post})$ = $\nabla_{n, n-1} \circ \nabla_{n-1, n-2} \circ \nabla_{n-2, n-3} \circ \dots \circ \nabla_{1, 0}$ = $\{\Omega_{in1}, \Omega_{in2}, \dots\}$

\begin{figure*}[t]
    \centering
  \begin{subfigure}[t]{0.45\textwidth}
  \begin{minipage}{0.80\textwidth}
  \begin{lstlisting}[language = C]
a = a->next;         
a = curr;            
if (curr != NULL){   
  b = b->next;       
  b = curr;          
  curr = curr->next; 
}                    \end{lstlisting}
 \end{minipage}
      \caption{Input program}
 \label{fig:reverse_1}
\end{subfigure}%
      ~ 
\begin{subfigure}[t]{0.55\textwidth}
      \centering
      \begin{minipage}{0.90\textwidth}
        \begin{lstlisting}[language = C]
(true) : a = a.next         
(true) : a = curr           
(true) : b0 = (curr != NULL)
(b0)   : b = b.next         
(b0)   : b = curr           
(b0)   : curr = curr.next   
\end{lstlisting}
  \end{minipage}
  \caption{IR of the program shown in \fig{fig:reverse_1}}
  \label{fig:reverse_2}  
  \end{subfigure}
 \caption{IR for a program with multiple statements in conditional block.}
\label{fig:no_name}
\end{figure*}

        

        
\begin{figure*}[t]
        \includegraphics[width=1.3\textwidth]{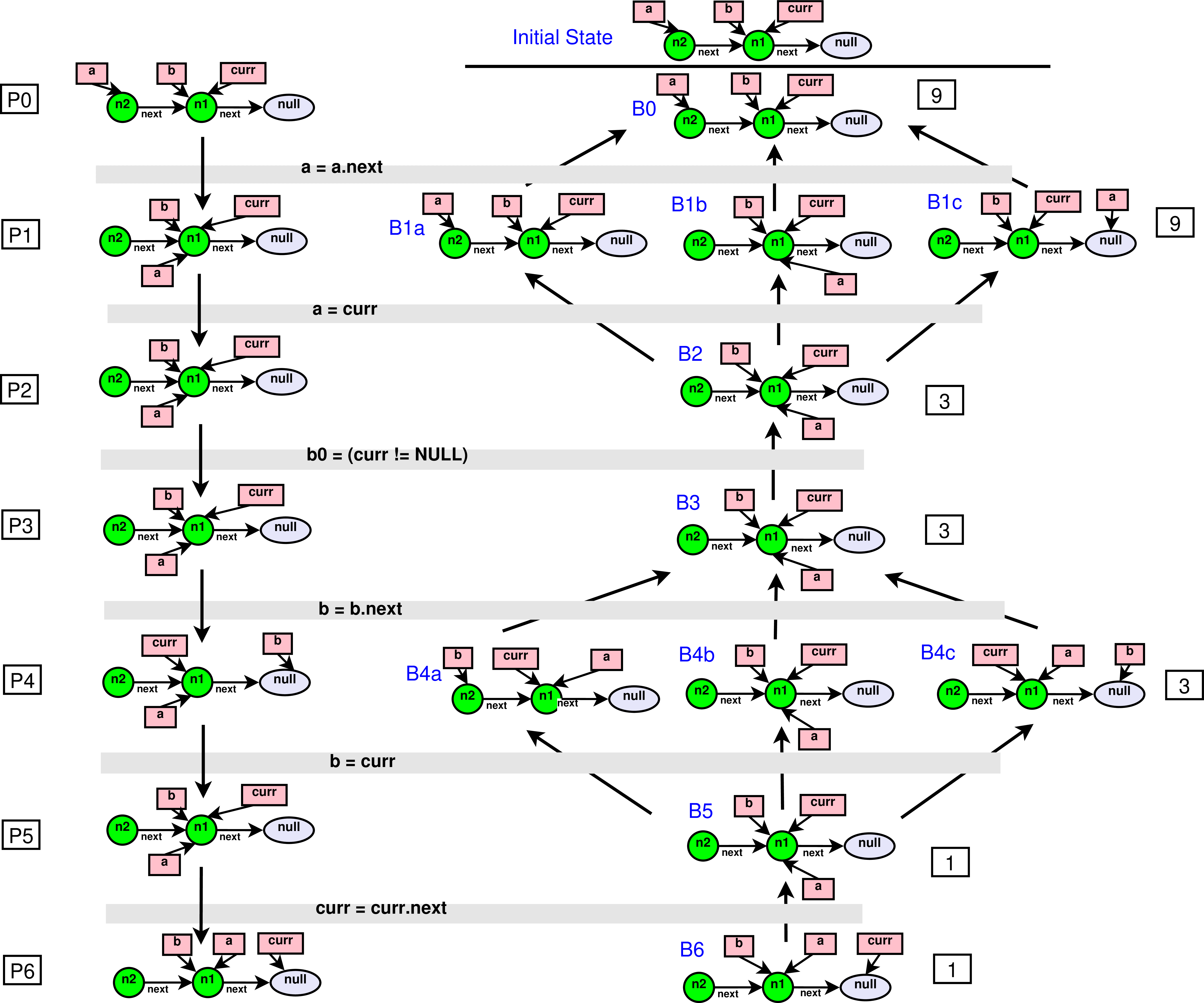}
        \caption{Forward and Backward execution of IR code shown in \fig{fig:reverse_2}}
        \label{fig:full_tree}
\end{figure*}

We define the backward execution in \fig{fig:full_tree}. Let us illustrate it using the program shown in \fig{fig:reverse_1} (and its IR in \fig{fig:reverse_2}). 
This program has no bug, and we only use it for elucidation.
Backward execution uses the rules described in \fig{fig:back_semantics}.
\textsf{P0} represents the program state before the start of the execution, and \textsf{P6} shows the program state after all statements have been executed.
We have copied the state \textsf{P0} to \textsf{Initial State} and state \textsf{P6} to \textsf{B6} on the right-hand side for convenience. 
$N$ is the set of all nodes (including \texttt{null}) in the program at a particular program point.
Starting from state \textsf{B6}, it uses \texttt{getfld\textsubscript{1}} rule, (because the LHS of statement $curr = curr\rightarrow next$ does not appear in any previous program statement) and this updates the value of \var{curr} to its value in the \textsf{Initial State} (node \var{n1}), as shown in state \textsf{B5}. 
We define two program statements as matching if their LHS writes to the same program variable or to a field of the same program variable. 
Now, since the LHS of statement $b = curr$ is same as LHS of previous statement $b = b\rightarrow next$, therefore we cannot read the value of \var{b} from the initial state. 
In this case, \var{b} is the variable to which both of these statements are writing.
The value of \var{b} has changed (from that in initial state \textsf{P0}) due to the presence of a matching statement. 
Since we cannot be sure of the actual value of \var{b} at this program point, to perform backward execution, we assign all possible node values (including \var{null}) to \var{b}. 
We call this process ``splitting of states". This is the use of \var{asgn\textsubscript{2}} rule and the set of possible states is \textsf{B4(a), B4(b) and B4(c)}.

For backward executing $b = b.next$, \toolname uses \var{getfld\textsubscript{1}} rule to update the value of \var{b}. 
This rule was used because, now, no previous program statement has the same LHS (\var{b}). 
Therefore, we update all the states (\textsf{B4(a), B4(b) and B4(c)}) with the value of \var{b} in the initial state (node \var{n1}) to give state \textsf{B3}. 
Although we maintain all the split states, we only show unique states in the figure.
Next, the \var{bassign} rule is used, which performs no updates to the state (shown in state \textsf{B2}). 
Now, we will again have to split states as the LHS of statement $a = curr$ and $a = a\rightarrow next$ are the same. 
\toolname uses the \var{asgn\textsubscript{2}} rule to assign all possible node values to \var{a}, and the obtained set of states are shown in \textsf{B1(a), B1(b) and B1(c)}. 
Finally, the \var{getfld\textsubscript{1}} rule is used to revise the value of \var{a}, and this time, it can be read from the initial state (node \var{n2}); this state is labelled as \textsf{B0}. 
Note that, in this case, state \textsf{B0} converges to the \textsf{Initial State}, showing that the program does not have a bug.

\vspace{5pt}

\begin{figure*}[t]
    \scalebox{0.8}{
    \begin{minipage}{\textwidth}
        \begin{align*}
                        \infer[getfld_{1}]
                        {\opsemb{\langle \mathcal{V}_1, \mathcal{H}_1 \rangle}{x\asgn y.f}{\langle \mathcal{V}_2, \mathcal{H}_1 \rangle}}
            {\Upsilon(x)=\Upsilon(y) & \mathcal{V}_2=\mathcal{V}_1[x\mapsto \mathcal{V}_{pre}(x)]\\ & \textrm{x:=y.f is upward exposed}}\hspace{70pt}
                        \infer[asgn_{1}]
                        {\opsemb{\langle \mathcal{V}_1, \mathcal{H}_1 \rangle}{x\asgn y}{\langle \mathcal{V}_2, \mathcal{H}_1 \rangle}}
            {\Upsilon(x)=\Upsilon(y) & \mathcal{V}_2=\mathcal{V}_1[x\mapsto \mathcal{V}_{pre}(x)]\\ & \textrm{x:=y is upward exposed}}
        \end{align*}

        \begin{align*}
                        \infer[getfld_{2}]
                        {\opsemb{\langle \mathcal{V}_1, \mathcal{H}_1 \rangle}{x\asgn y.f}{\langle \mathcal{V}_2, \mathcal{H}_1 \rangle}}
            {\Upsilon(x)=\Upsilon(y.f) & \mathcal{V}_2=\mathcal{V}_1[x\mapsto n \ | \ n \ \overset{*}{\gets} \ N]\\ & \textrm{x:=y.f is not upward exposed}}  \hspace{70pt}
                        \infer[asgn_{2}]
                        {\opsemb{\langle \mathcal{V}_1, \mathcal{H}_1 \rangle}{x\asgn y}{\langle \mathcal{V}_2, \mathcal{H}_1 \rangle}}
                        {\Upsilon(x)=\Upsilon(y) & \mathcal{V}_2=\mathcal{V}_1[x\mapsto n \ | \ n \ \overset{*}{\gets} \ N]\\ & \textrm{x:=y is not upward exposed}}
        \end{align*}
        
        \begin{align*}
                        \infer[putfld_{1}]
                        {\opsemb{\langle \mathcal{V}_1, \mathcal{H}_1 \rangle}{x.f \asgn y}{\langle \mathcal{V}_1, \mathcal{H}_2 \rangle}}
                        {\Upsilon(x.f)=\Upsilon(y) & \mathcal{H}_2=\mathcal{H}_1[(\langle \mathcal{V}_p(x), f \rangle) \mapsto \mathcal{H}_p(\langle \mathcal{V}_p(x), f \rangle)]\\ & \textrm{stmt is upward exposed}}\hspace{20pt}
                        \infer[grd]
                        {\opsemb{\langle \mathcal{V}_1, \mathcal{H}_1\rangle}{grd\text{ ? }stmt}{\langle \mathcal{V}_2, \mathcal{H}_2\rangle}}
                        {\opsemb{\langle \mathcal{V}_1, \mathcal{H}_1\rangle}{\text{stmt}}{\langle \mathcal{V}_2, \mathcal{H}_2\rangle}}
        \end{align*}

        \begin{align*}    
                        \infer[putfld_{2}]
                        {\opsemb{\langle \mathcal{V}_1, \mathcal{H}_1 \rangle}{x.f \asgn y}{\langle \mathcal{V}_1, \mathcal{H}_2 \rangle}}
                        {\Upsilon(x.f)=\Upsilon(y) & \mathcal{H}_2=\mathcal{H}_1[(\langle \mathcal{V}_1(x), f \rangle) \mapsto n \ | \ \forall \ n \ \epsilon \ N]\\ & \textrm{stmt is not upward exposed}}\hspace{20pt}
                        \infer[noop]
                        {\opsemb{\langle \mathcal{V}_1, \mathcal{H}_1\rangle}{command}{\langle \mathcal{V}_1, \mathcal{H}_1\rangle}}
                                                {command \in \{band, bor, bassign, skip\}}
        \end{align*}


    \end{minipage}
        }
                \caption{\textbf{Backward Semantics} of our intermediate representation}
    \label{fig:back_semantics}
\end{figure*}

\subsubsection{Bug Localization Algorithm}

We start our discussion about the algorithm by stating the following lemmas.  Let $\Omega$ denote a set of states, while $\omega$ represents a single state.
    
    \begin{lemma}
        If $i^{th}$ statement is upward exposed then,
        $\Delta_{0, i-1}(\omega_{pre}, P)[LHS(s_i)]\ = \ \omega_{pre}[LHS(s_i)]$.
        \label{lemma_basic1}
    \end{lemma}
    \begin{proof}
        Since $i^{th}$ statement is upward exposed, no other preceding statement or aliased variable can change the value of $LHS(s_i)$, hence its value remains the same as in precondition ($\omega_{pre}$).
    \end{proof}

    \begin{lemma}
        If $i^{th}$ statement is downward exposed then, 
        $\nabla_{n, i+1}(\omega_{post}, P)[LHS(s_i)]\ = \ \omega_{post}[LHS(s_i)]$.
        \label{lemma_basic2}
    \end{lemma}
    \begin{proof}
        Since $i^{th}$ statement is downward exposed, no other succeeding statement or alias variable changes the value of $LHS(s_i)$, hence its value remains the same as in postcondition ($\omega_{post}$) during backward execution. There has not been any splitting of states for $LHS(s_i)$ before $i^{th}$ statement is backward traversed.
    \end{proof}

We provide our complete bug localization algorithm (\alg{alg:buglocalization}) for detecting suspicious program statements. 
Line~2 in~\alg{alg:buglocalization} is the forward trace, that is, the sequence of states obtained from the forward execution of the program. 
Line~3 is the backward trace, or the sequence of states obtained from backward execution; as we have seen, spitting can happen in the backward execution leading to several states at a program point (so $\Omega_i$ is a set of states). 
Since the splitting is uniform, we can count the number of children of a particular state in the backward trace by dividing the final number of states at the end of backward execution (denoted by $\var{M}$) and the number of states at that program point. 
This process gives a multiplying factor at each program point. 
Line~4 calculates the distance between the state in forward execution and the state(s) in backward execution, multiplied by the multiplying factor pertaining to that point. 
This operation is denoted by $\otimes$ in the algorithm. 
Each $\Gamma_i$ stores the result of this operation.
Starting from the last program statement, if the pairwise difference of the $\Gamma_i$'s is non-zero (this is the gradient), the $i^{th}$ statement is added to the set of suspicious statements (Line~5-7), which is returned at the end (Line~9).


Let the number of states in ${\nabla(P, \Omega_{post}, i)}$ be $\var{p}$ and the number of states at the end of backward execution ${\nabla_{n, 0}(P, \Omega_{post})}$ be $\var{M}$.

Let $\nabla_{n, i} = \Omega^b_i$ and $\Delta_{1, i} = \omega^f_i$. Then,

\[ \Gamma_i = \omega^f_i \otimes \Omega^b_i = \frac{M}{\vert \Omega^b_i \vert} * ( \vert\Omega^b_i[n] - \omega^f_i\vert + \vert\omega^f_i[n] - \Omega^b_i\vert ) \]    

In \fig{fig:full_tree}, upon complete backward execution, we get a total of 9 states. 
The number of states at every program point is shown on the right-hand side (in the aligned box) \fig{fig:full_tree}.
The multiplying factor for a state can be determined by dividing 9 by its number of states. 


$\Gamma_1$ at P1 is the sum of the difference between the nine states in backward execution (right side), out of which only three are shown (as others are duplicated). With the state in forward execution (left side) at P1, the sum of differences is \texttt{6}.

We provide the theoretical analysis of this algorithm in the Appendix.

\RestyleAlgo{boxruled}
\begin{algorithm}[t]
        \KwIn {$\phi :: [\langle P, n, \Omega_{pre}, \Omega_{post}\rangle]$}
                \caption{\label{alg:buglocalization}Pick suspicious statements}
                $SuspiciousSet = \{\}$ \\
        $[\omega^f_0, \omega^f_1, \omega^f_2, \dots, \omega^f_n] \gets [\mathcal{I}, \Delta_{1, 1}, \Delta_{1, 2}, \dots \Delta_{1, n}] (\omega_{pre}$, P) \\ 
        $[\Omega^b_0, \Omega^b_1, \Omega^b_2, \dots, \Omega^b_n] \gets [\nabla_{n, 1}, \nabla_{n, 2}, \dots \nabla_{n, n}, \mathcal{I}] (\omega_{post}$, P) \\ 
                $[\Gamma_0, \Gamma_1, \Gamma_2, \dots, \Gamma_n] \gets [\omega^f_0, \omega^f_1, \omega^f_2, \dots, \omega^f_n] \otimes [\Omega^b_0, \Omega^b_1, \Omega^b_2, \dots, \Omega^b_n]$ \\ 
                \For{$i\in \{1 \dots n\}$}{
        \uIf{$\Gamma_{i-1} \neq \Gamma_{i}$}    
                {$SuspiciousSet = SuspiciousSet \cup  \{i\}$}   
                }
                \Return SuspiciousSet

\end{algorithm}

\section{Advanced Debugging/Repair}
\label{sec:advanced}

In this section, we show how skilled engineer can employ the features in \toolname for effective debug-repair sessions.

\subsection{Specification Refinement}
\label{sec:spec_refine}
\toolname is designed to model heap manipulations; however, \toolname can use the \var{concrete($\zeta$)} statement in its intermediate representation as an abstraction of any statement $\zeta$ that it does not model. 
On hitting a \var{concrete($\zeta$)} statement, \toolname uses \var{gdb} to concretely execute the statement and updates its symbolic state from the concrete states provided by \var{gdb}. 
\fig{fig:refine2} shows an instance where we wrap the \var{i=i+1} statement in an concrete execution; \toolname translates this statement to a string of \var{gdb} commands, and the symbolic state is updated with the value of $i$ from the concrete state that \var{gdb} returns after executing the statement. 
Hence, although \toolname is specifically targeted at heap manipulations, it can also be used to debug/repair programs containing other constructs as long as the bug is in heap manipulation statements. 
We refer to this technique of reconstructing the symbolic specification by running the statement concretely as {\it specification refinement}.

{\it Specification refinement} can be used in creative ways by skilled engineers. 
In \fig{fig:refine2}, the programmer decided to wrap a complete function call (\var{foo()}) within the \var{concrete()} construct, allowing \toolname to reconstruct the effect of the function call via concrete execution without having to model it. 
This strategy can fetch significant speedups for repair: let us assume that, in \fig{fig:motivating}, the programmer uses her domain knowledge to localize the fault to Lines~6--8; she can pass this information to \toolname by wrapping the other statements in the loop (lines~5,9) in \var{concrete} statements; this hint brings down the repair time on the full program on the complete execution from 6.0s to 1.5s, i.e., achieving a 4$\times$ speedup (on our machine). 
For this experiment, we turned off the fault localizer in \toolname.
This is understandable as each instruction that is modeled can increase the search space exponentially.

\subsection{Checkpoint-based Hopping}
\label{sec:checkpoint-hop}

\begin{figure}[t]
\centering
\begin{minipage}{\columnwidth}
\begin{lstlisting}[language = C]
void bar(){
     struct node *current=NULL; int i;
     current = head;
     while (current != NULL){
         concrete[i = foo();] //concrete stmt
         current->data = i; 
         current = current->next;
     }
}
\end{lstlisting}
\caption{Refinement with concrete function calls}
\label{fig:refine2}
\end{minipage}%
\hfill
\begin{minipage}{\columnwidth}
\begin{lstlisting}[language = C]
struct node *head;
void reverse(){
     struct node *current, *temp1=NULL, *temp2=NULL;
     current = head; 
     while (current != NULL){ // CheckPt 0
         temp1 = current->prev;
         temp2 = current->next;
         current->prev = temp1; // FIX1: LHS = temp2;
         current->next = temp2; // FIX2: LHS = temp1;
         current = current->prev; 
     } // CheckPt 1; CheckPt 2; CheckPt 3
     head = temp1->prev;
}
...
int main(){
 push(2); push(4); push(8); push(10);
 reverse();
}

\end{lstlisting}
\caption{Example for checkpoint-based hopping}
\label{fig:tree_debugging}
\end{minipage}
\end{figure}

\begin{figure*}
 \centering
 \begin{subfigure}[b]{0.5\textwidth}
 \centering
 \includegraphics[width=\textwidth]{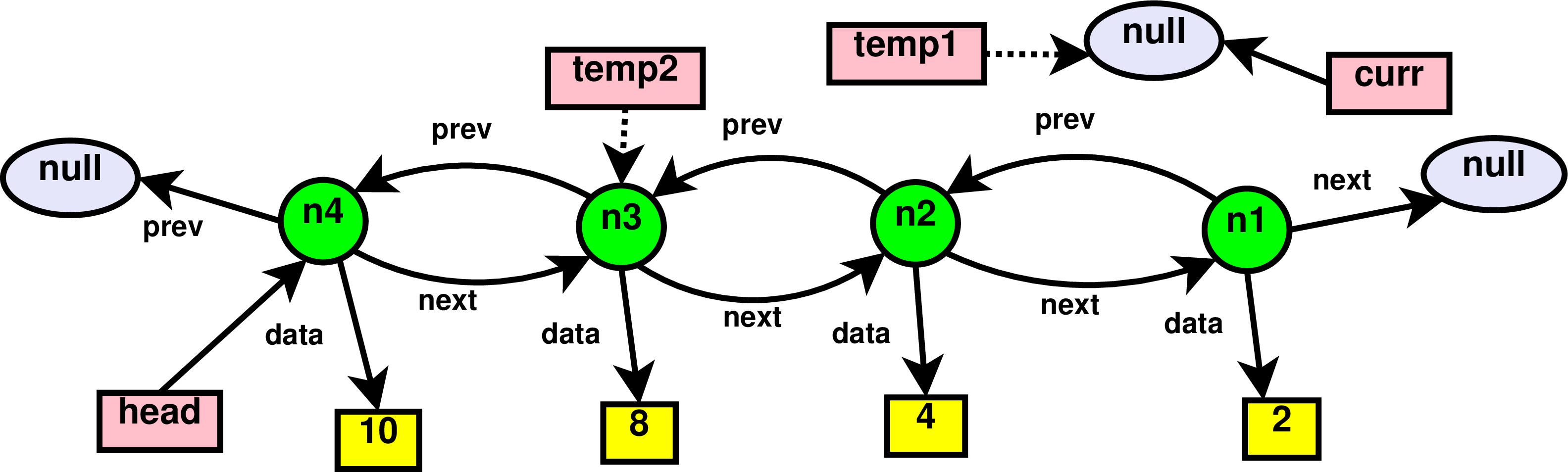}
 \caption{At the end of first loop iteration \\ (before user changes)}
 \label{fig:first_TD}
 \end{subfigure}%
 \hfill
 \begin{subfigure}[b]{0.5\textwidth}
 \centering
 \includegraphics[width=\textwidth]{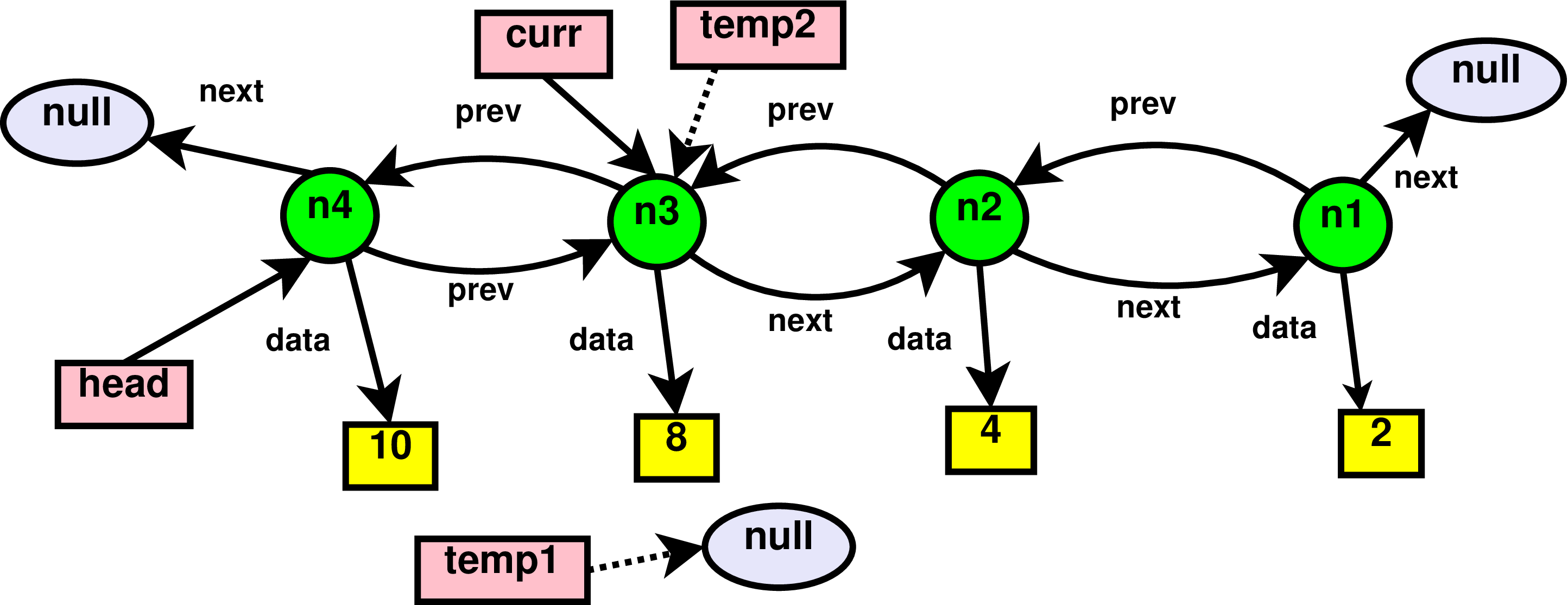}
 \caption{At the end of first loop iteration \\ (after user changes)}
 \label{fig:first_change_TD}
 \end{subfigure}%
 \\
 \begin{subfigure}[b]{0.5\textwidth}
 \centering
 \includegraphics[width=\textwidth]{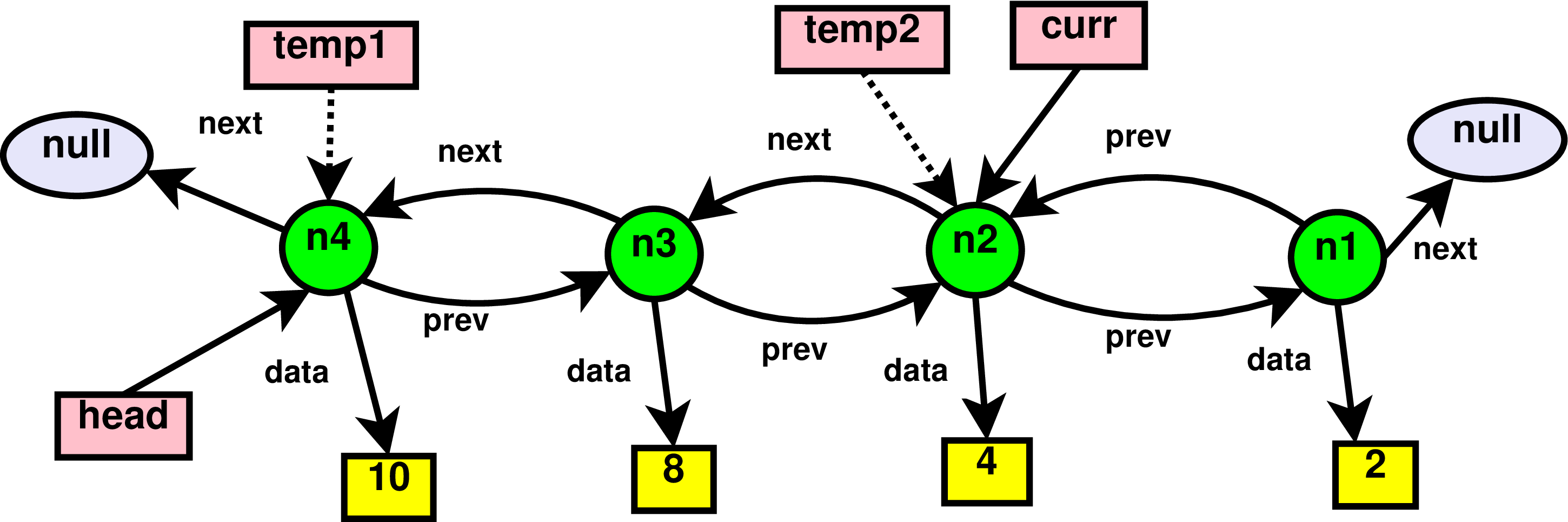}
 \caption{At the end of second loop iteration \\ (after user changes)}
 \label{fig:second_TD}
 \end{subfigure}%
 \hfill
 \begin{subfigure}[b]{0.5\textwidth}
 \centering
 \includegraphics[width=\textwidth]{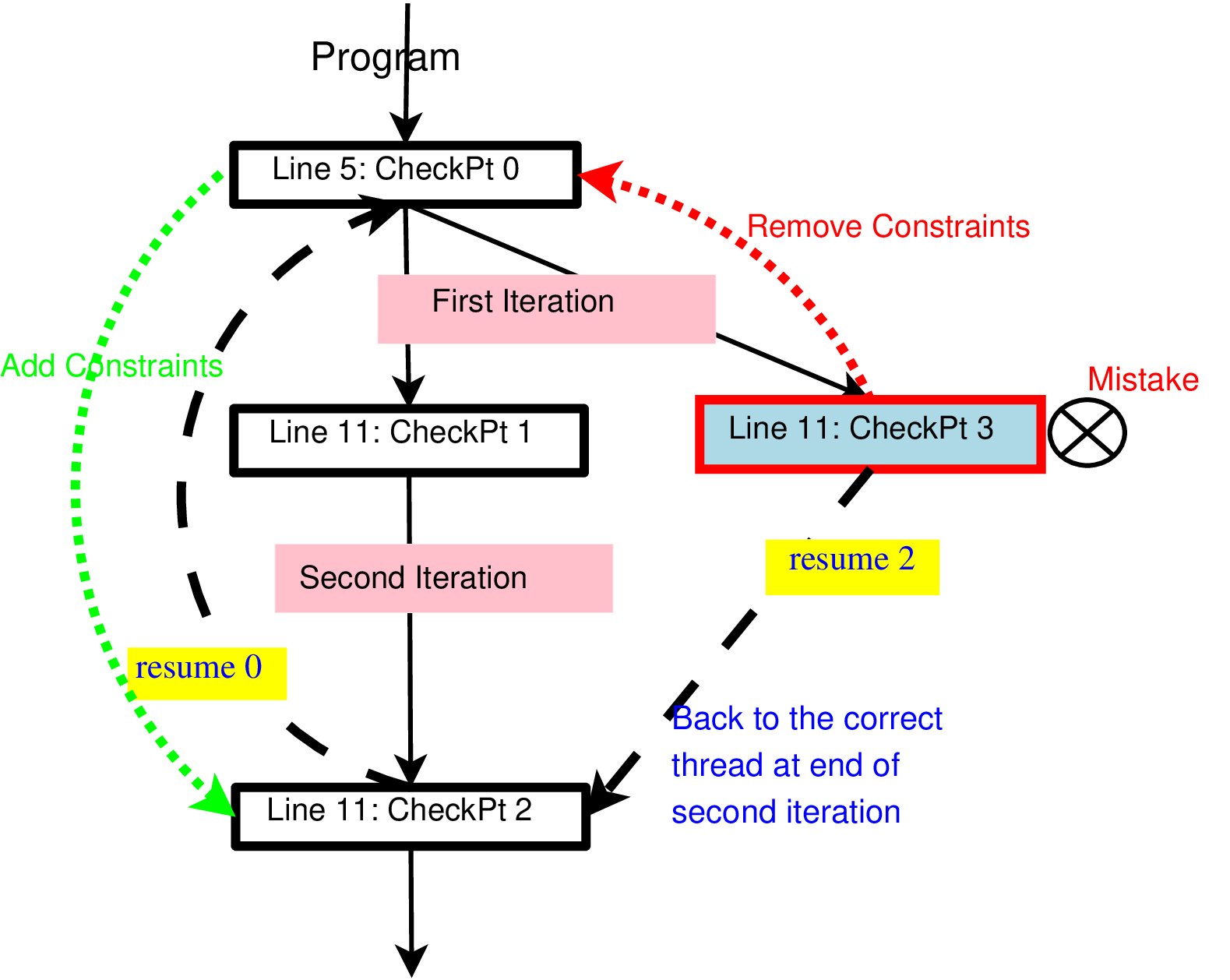}
 \caption{Graph showing initial and final checkpoints for \colorbox{yellow}{resume0} and \colorbox{yellow}{resume2}, and handling of the constraints for \colorbox{yellow}{resume2}.}
 \label{fig:tree_TD}
 \end{subfigure}%
 \caption{Figures for checkpoint-based hopping example}
 \vspace{-2pt}
\end{figure*}

This feature comes in handy when the programmer herself is not sure about the correctness of the expected specification she is asserting. 
For example, \fig{fig:tree_debugging} shows the code for reversal of a doubly-linked list, with bugs in two statements in the \var{while} loop (these bugs are different from the ones in \fig{fig:motivating}).


The programmer steps into the \var{reverse()} function after creating the doubly linked-list.

\colbox{{\prmt start\\ Starting program...\\ push(2)\\\prmt next; next; next; next;}}

\colbox{{push(4);\\\dots}}

\colbox{{reverse();\\\prmt step}}

\colbox{{current = head; \\ \prmt next}}

\noindent She executes the first statement in \var{reverse()} and then asserts the program state. Whenever the user asserts a state, \toolname creates a checkpoint of this state at that program point; checkpointing memorizes important events during a debug run, allowing the user to resume a new direction of debugging from this location, if required (illustrated later). 
A \textit{Checkpoint ID}, which keeps a count of the checkpoints (0 in this case), is returned. 
This ID can be used to resume debugging from this corresponding checkpoint/

\colbox{{while(current != NULL) \\ \prmt spec\\ Program states added -- Checkpoint 0\\ \prmt enter}}

She employs the \underline{{\tt next}} command to execute till the end of the loop.

\colbox{{temp1 = current->prev;\\ \prmt next; next; next; next; next;\\ \dots \\ while(current != NULL)}}

\noindent The program state displayed to her is shown in \fig{fig:first_TD}: due to the fact that the effect of Bug\#1 and Bug\#2 cancel out, she observes that the data-structure has not changed except the \var{current} pointer. 
Since she expected the reversal of the first node by the end of this loop iteration, she issues the desired changes and asserts the state (Checkpoint 1); the updated heap is shown in \fig{fig:first_change_TD}.

\colbox{\prmt change n4 -> prev n3\\ \prmt change current -> n3}\\
\colbox{{while(current != NULL) \\ \prmt spec\\ Program states added -- Checkpoint 1}}


\noindent She now enters the loop for the second time.\\
\colbox{{\prmt enter\\ while(current != NULL)\\ \dots}}

\noindent At the end of this iteration, the user again finds an unexpected state and issues necessary changes to reverse the next node.

\colbox{\prmt change current n2 \\ \prmt change n3 -> prev n2\\ \prmt change n3 -> next n4}

\noindent Satisfied with the updated state (shown in \fig{fig:second_TD}), she asserts it (Checkpoint 2).

\colbox{{while(current != NULL) \\ \prmt spec\\ Program states added -- Checkpoint 2}}

\noindent She, now, feels less convinced about her hypothesis regarding the correct run of the program and, thus, about the asserted program states 
and, therefore, decides to try out another direction of investigation. 
This would have required her to abandon the current session and spawn a new session; not only will it require her to resume debugging from the beginning, 
but she would also lose the current debugging session, preventing her from resuming in case she changes her mind again.
\toolname packages feature for such a scenario where a programmer may be interested in exploring multiples directions, allowing them to save and restore among these sessions at will. 
In this case, instead of exploring from the initial state, the programmer adds a checkpoint for the current state (Checkpoint 2) and returns to \var{Checkpoint 0} by issuing \underline{\var{resume 0}}. 

\colbox{{while(current != NULL) \\ \prmt resume 0\\ Program resumed at checkpoint 0}}

She executes through the first loop iteration in a similar manner (thus, the obtained state is the same as shown in \fig{fig:first_TD}). She issues the following changes to match her expectations and asserts the updated program state (Checkpoint 3).

\colbox{\prmt change current n3 \\ \prmt change n3 -> prev n4 \\ \prmt change n3 -> next null}\\
\colbox{{while(current != NULL) \\ \prmt spec\\ Program states added -- Checkpoint 3}}

To her surprise, the state shown to her remains unchanged after these modifications (barring the \var{current} pointer). She is now more confident that the states she had asserted in the previous session were correct and now wants to return to it. She issues \underline{\var{resume 2}}.

\toolname essentially maintains the constraints corresponding to the checkpointed states along with the different debugging states in a directed-acyclic graph; when the programmer resumes from checkpoint 2, \toolname pops off all the constraints that were asserted till the first common ancestor (Checkpoint 0) of the current (Checkpoint 3) and the requested checkpointed state (Checkpoint 2) from the solver, and then, pushes the constraints till the resumed checkpoint~(Checkpoint 2). 
They have been shown as red and green dotted lines, respectively, in \fig{fig:tree_TD}.

\section{Experiments}
\label{sec:experiments}

\begin{table*}
        \centering
        \caption{Our two sets of benchmarks\protect\cite{GeeksForGeeks}}
        \scalebox{1.0}{
        \begin{tabularx}{\textwidth}{| c | X || c | X |}
        \multicolumn{4}{c}{\textbf{Small benchmarks}} \\ \toprule
        B1 & Reverse singly linked-list & B2 & Reverse doubly linked-list \\ 
        B3 & Deletion from singly linked-list & B4 & Creation of circular linked-list \\
        B5 & Sorted Insertion singly linked list & B6 & Insertion in single linked list \\ 
        B7 & Swapping nodes singly linked list & B8 & Splaytree Left Rotation \\
        B9 & Minimum in Binary Search Tree & B10 & Find Length of singly linked list \\
        B11 & Print all nodes singly linked list & B12 & Splitting of circular linked list\\ 
        B13 & AVL tree right rotation & B14 & AVL tree left-right rotation \\
        B15 & AVL tree left rotation & B16 & AVL tree right-left rotation \\
        B17 & Red-Black tree left rotate & B18 & Red-Black tree right rotate\\
        B19 & Enqueue using linked-list & B20 & Splaytree Right Rotated \\
        \bottomrule
        \multicolumn{4}{c}{\textbf{Large benchmarks}} \\
        \toprule
        L1 & Delete middle of singly linked-list &  L2 & Remove dupilcate in singly linked-list \\
        L3 & Last node to first singly linked-list & L4 & Intersection of two singly linked-list \\
        L5 & Split singly linked-list into two lists & L6 & Value-based partition of singly linked-list \\ 
        L7 & Delete specific node in singly linked-list & L8 & Splitting of circular linked-list  \\
        L9 & Middle node as head singly linked-list & L10 & Merge alternate nodes in singly linked-list \\
        L11 & Delete node of specific value in singly linked-list & L12 & Splitting of doubly linked-list \\ 
        L13 & Pairwise swap of nodes in singly linked-list &  L14 & Rearranging singly linked-list \\
        L15 & Absolute sort of nodes in singly linked-list & L16 & Quicksort of singly linked-list \\
        L17 & Delete specific node in doubly linked-list & L18 & Sorted insertion in singly linked-list \\ 
        L19 & Remove duplicate nodes in doubly linked-list & L20 & Constrained deletion in singly linked-list  \\ 
        \bottomrule
        \end{tabularx}
        }
        \label{tab:all_benchmarks}
\end{table*}

We built \toolname using the \var{gdb} Python bindings\cite{gdbapi}, the C-to-AST compiler uses \var{pycparser}\cite{pycparser}, the visualization module uses \var{igraph}\cite{igraph} to construct the box and arrow diagrams and the repair module uses the Z3\cite{z3} theorem prover to solve the SMT constraints. 
We conduct our experiments on an Intel(R) Xeon(R) CPU @ 2.00GHz machine with 32~GB RAM. 
To evaluate our implementation, we attempt to answer the following research questions:
\begin{description}
 \item[{\bf RQ1.}] Is our repair algorithm able to fix different types and combinations of bugs in a variety of data-structures?
 \item[{\bf RQ2.}] Can our repair algorithm fix these bugs in a reasonable time?
 \item[{\bf RQ3.}] How does our repair algorithm scale as the number of bugs is increased? %
 \item[{\bf RQ4.}] What is the accuracy of our localization algorithm with respect to other localization algorithms? 
 \item[{\bf RQ5.}] What is the impact of our algorithm (localization + repair) on the repair time of \toolname? 
 \item[{\bf RQ6.}] Is \toolname capable of debugging/fixing real bugs? 
\end{description}



We conduct our study on 40 heap manipulating programs~(\tab{tab:all_benchmarks}) from online sources~\cite{GeeksForGeeks} for a variety of data-structures like singly, doubly, and circular linked lists, AVL trees, Red-Black trees, Splay Trees, and Binary Search Trees. 

Though there has been a large body of work on automated debugging and repair~\cite{anagelicDebugging:2011,Angelix:2016,prophet:2016,genProg:2012,semfix:2013,Modi:2013,nguyen:2009,weimer:2009,weimer:2006,Loncaric:2018,Abhik:2016,Nguyen:2019,Ulysis,Kolahal,Gambit,Pandey:2019,Bavishi2016b}, these techniques cannot tackle repair over deep properties like functional correctness of heap data-structures. 
Our work is more in line with the following papers involving synthesis and repair of heap manipulating programs~\cite{Singh:2011,similarbenchmark1,similarbenchmark2,Roy:2013,Garg:2015,similarbenchmark3,similarbenchmark12} or that involve functional correctness of student programs~\cite{similarbenchmark4,similarbenchmark5,similarbenchmark6,similarbenchmark7,similarbenchmark8,similarbenchmark9,similarbenchmark10,similarbenchmark11,similarbenchmark13}. 
Hence, our collection of benchmarks are similar to the above contributions. 

We divide our benchmarks into \texttt{Small} and \texttt{Large} benchmarks: the \texttt{Large} benchmarks involve more complex control-flow (nested conditions and complex boolean guards) and are about 3$\times$ larger than the \texttt{Small} programs (in terms of the number of IR instructions).

\subsection{Experiments with Fault-Injection}
We create buggy versions via an in-house fault injection engine that automatically injects bugs (at random), thereby eliminating possibilities of human bias. 
For each program, we control our fault-injection engine to introduce a given number of bugs. 
We characterize a buggy version by $\langle x, y \rangle$, implying that the program requires mutation of $x$ (randomly selected) program expressions and the insertion of $y$ newly synthesized program statements.
The value of $x$ is determined by the number of mutations in the correct program. 
A mutation consists of replacing a program variable or the field of a variable with another variable or another field (randomly chosen from the program space) in a program statement or a guard.
The value of $y$ is determined by the number of program statements the engine deletes from the correct program. 
\toolname is unaware of the modifications and deletions when fed the modified program. \\

For the experiments, \toolname makes ten attempts at repairing a program, each attempt followed by \textit{proof-directed search space widening}; each attempt is run with a timeout of 30s. The experiment was conducted in the following manner:
\begin{enumerate}[noitemsep,wide=5pt, leftmargin=25pt]
 \item We evaluate each benchmark (in \tab{tab:all_benchmarks}) for four bug classes: Class1 ($\langle 1, 0\rangle$), Class2($\langle 1, 1\rangle$), Class3($\langle 2, 0\rangle$) and Class4($\langle 2, 1\rangle$);
 \item For each benchmark $B_i$, at each bug configuration $\langle x, y\rangle$, we run our fault injection engine to create \textit{40 buggy versions} with \var{x} errors that require modification of an IR instruction and \var{y} errors that require insertion of a new statement;
 \item Each of the above buggy programs is run twice to amortize the run time variability.
\end{enumerate}

\begin{figure}
        \centering
        \begin{subfigure}[b]{0.5\textwidth}
        \centering
        \includegraphics[width=\textwidth]{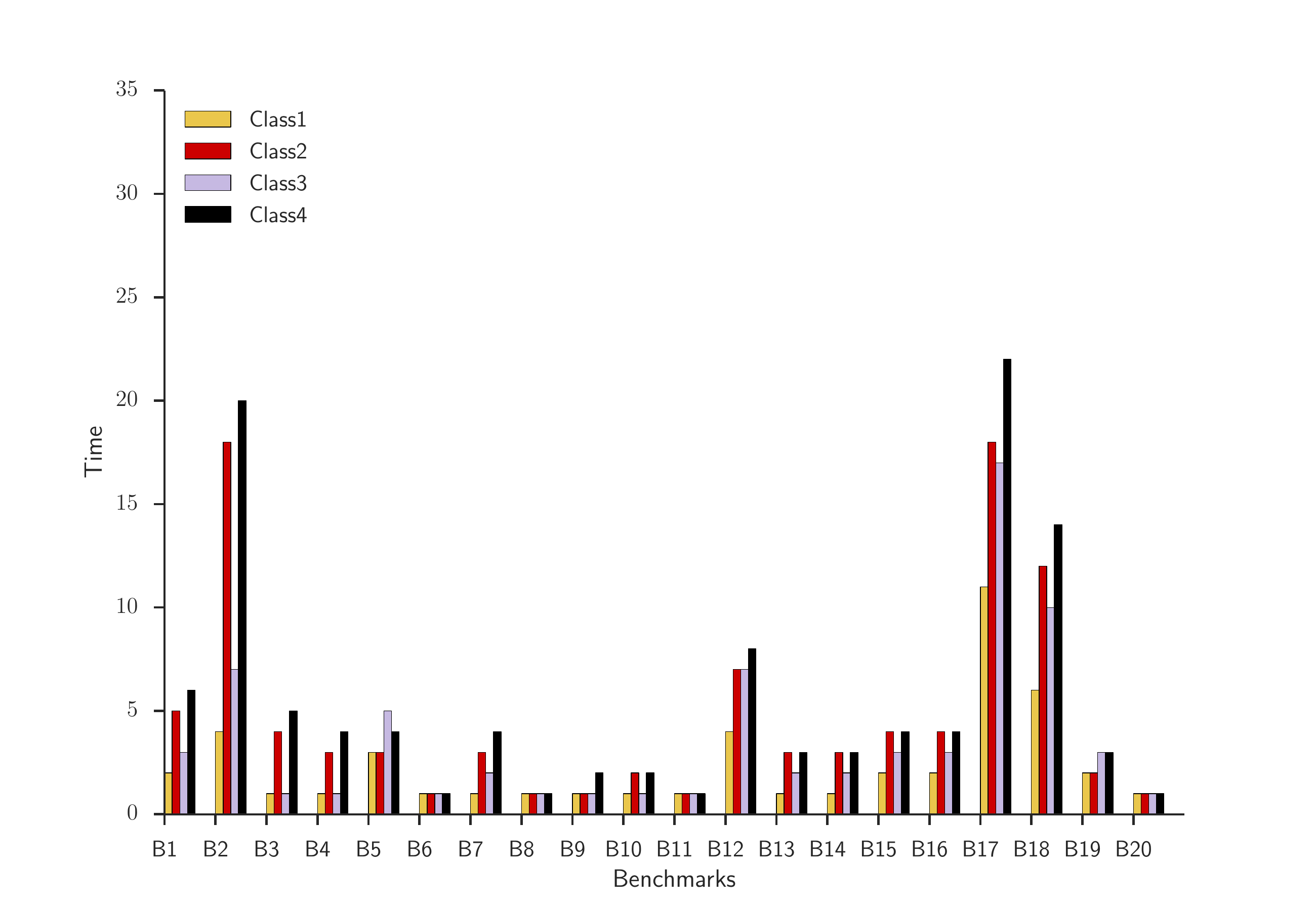}
        \caption{\vspace{-1pt}Repair time for our primary algorithm}
        \label{fig:algob2}
        \end{subfigure}%
        \hfill
        \begin{subfigure}[b]{0.5\textwidth}
        \centering
        \includegraphics[width=\textwidth]{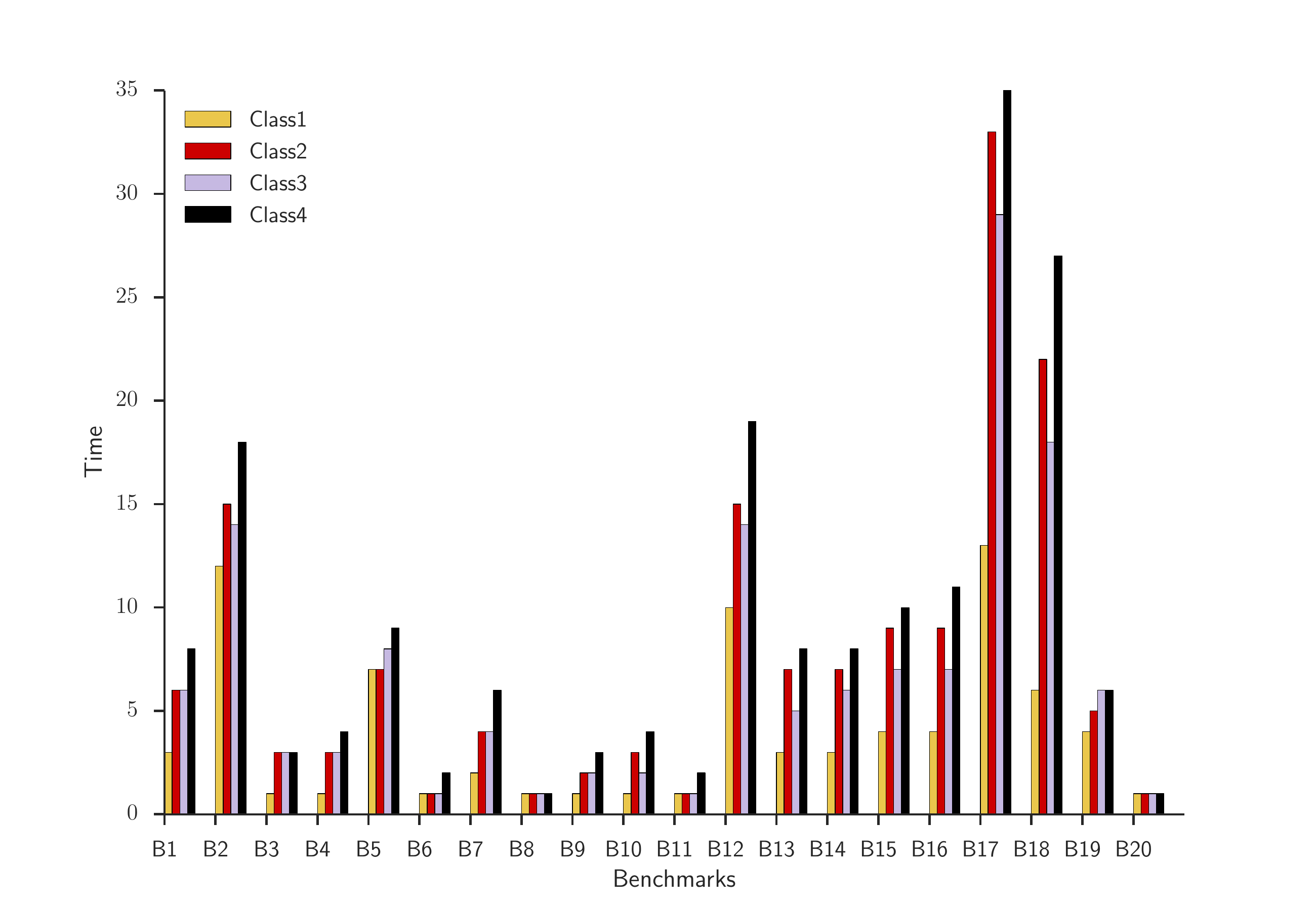}
        \caption{\vspace{-1pt}Repair time for the variant (AlgVar) \\ of our primary algorithm}
        \label{fig:algo4}
        \end{subfigure}%
        \\
        \begin{subfigure}[b]{0.5\textwidth}
        \centering
        \includegraphics[width=\textwidth]{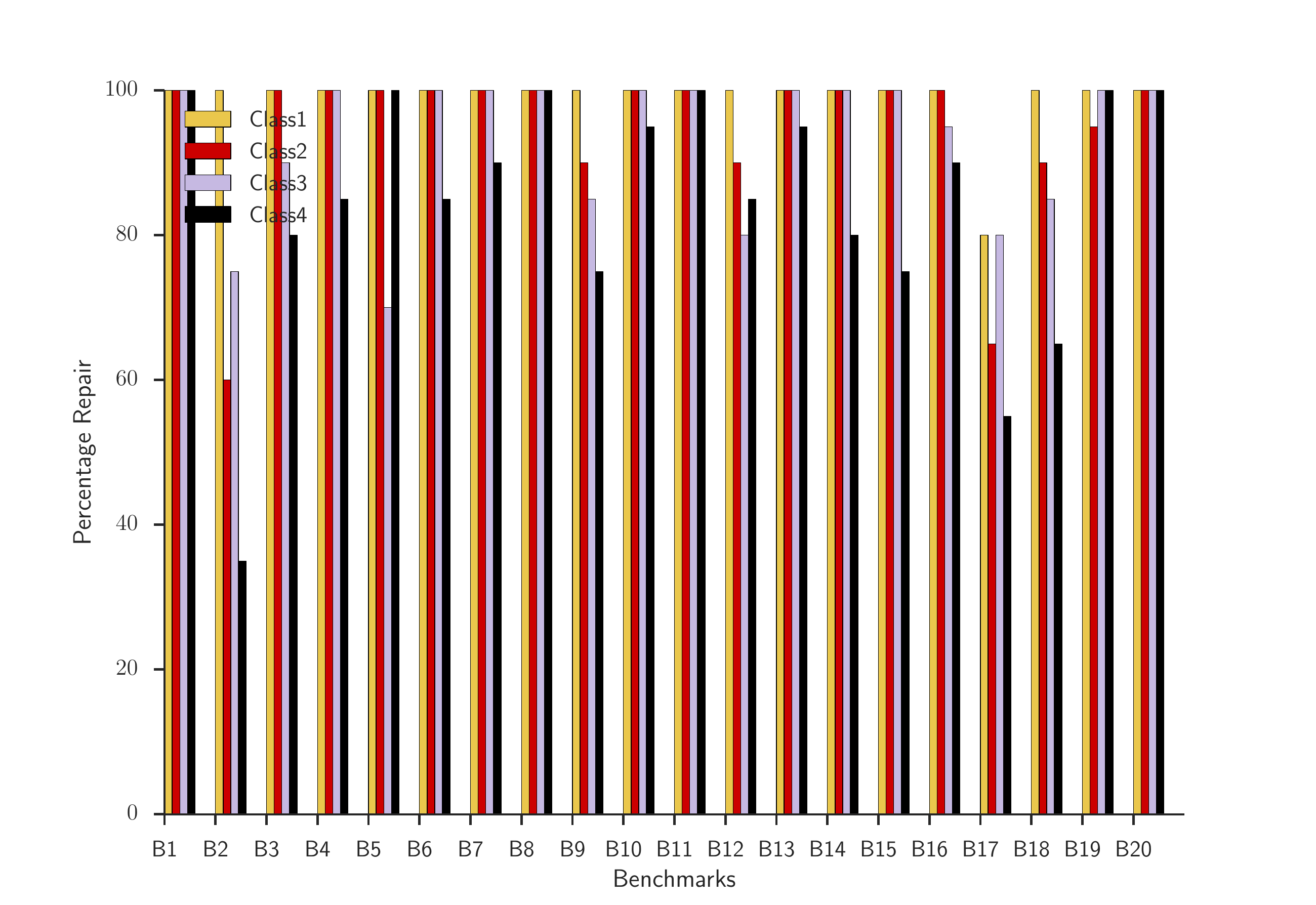}
        \caption{\vspace{-1pt}Success rates for AlgVar; our primary \\ algorithm has 100\% success rate in all cases.}
        \label{fig:repairplot}
        \end{subfigure}%
        \hfill
        \begin{subfigure}[b]{0.5\textwidth}
        \centering
        \includegraphics[width=\textwidth]{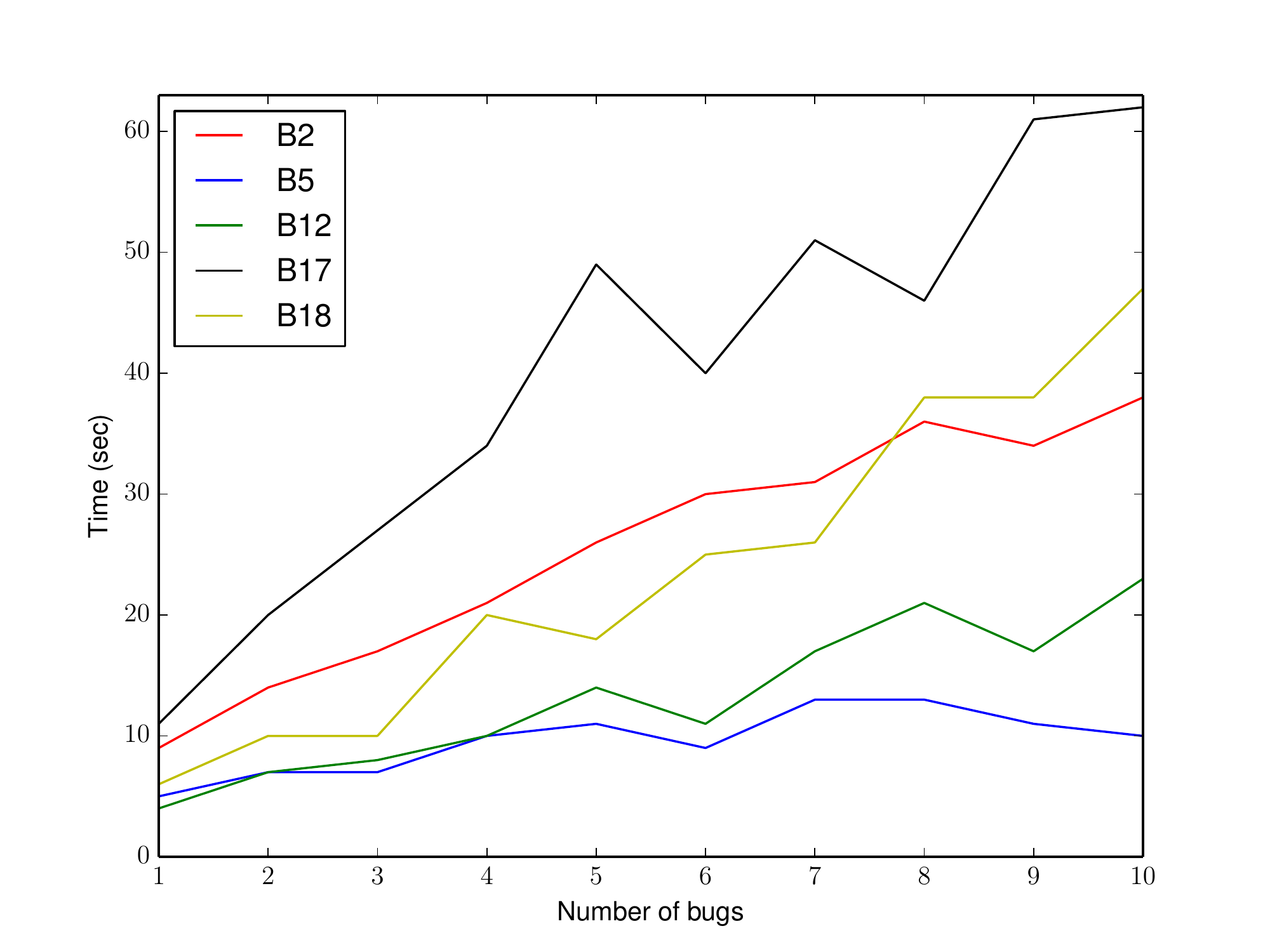}
        \caption{\vspace{-1pt}Experiment showing the increase in repair time with increasing number of bugs}
        \label{fig:increasing-bugs}
        \end{subfigure}%
        \caption{Performance of our repair algorithms}
        \label{fig:fse_overall}
\end{figure}

For RQ1 and RQ2, we use the 20 \texttt{Small} programs as these experiments involve only the repair tool (sans the localizer).
\fig{fig:fse_overall} shows the average time taken to repair a buggy configuration over the 20 buggy variants, which were themselves run twice (the reported time shows the average time taken for the successful repairs only). 
We report the time taken for our main algorithm (in \alg{alg:repair}) and its variant \textbf{AlgVar} (discussed in the last paragraph in \sect{sec:Repair}). 
Our primary algorithm performs quite well, fixing most of the repair instances in less than 5 seconds; understandably, the bug classes that require insertion of new instructions (Classes 2 and 4) take longer. 
There were about 1--4 widenings for bugs in class 1,2,3; the bugs in class 4 were more challenging, needing 2--6 widenings. 


In terms of the success rate, \textit{our primary algorithm was able to repair all the buggy instances}. However, \fig{fig:repairplot} shows the success rate for each bug configuration for \textbf{AlgVar}; the success rate is computed as the fraction of buggy instances (of the given buggy configuration) that could be repaired by \toolname (in any of the two attempts).

The inferior performance of the variant of our main algorithm shows that the quality of the unsat cores is generally poor, while the performance of our primary algorithm demonstrates that even these unsat cores can be used creatively to design an excellent algorithm. 

\fig{fig:increasing-bugs} answers RQ3 by demonstrating the scalability of \toolname with respect to the number of bugs on (randomly selected) five of our smaller benchmarks. We see that in most of the benchmarks, the time taken for repair grows somewhat linearly with the number of bugs, though (in theory), the search space grows exponentially. 
Also, one can see that more complex manipulations like left-rotation in a red-black tree (B17) are affected more as a larger number of bugs are introduced compared to simpler manipulations like inserting a node in a sorted linked list (B5).

The variance in the runtimes for the different buggy versions, even for those corresponding to the same buggy configuration, was found to be high. This is understandable as SMT solvers often find some instances much easier to solve than others, even when the size of the respective constraint systems is similar.

For RQ4, we compared our algorithm with the state-of-the-art bug localization algorithms we found in literature~\cite{Sober:2005,liblit:2005,Tarantula:2005,Oichai:2009}. 
All these algorithms require the creation of test cases that differentiate between the correct and incorrect behavior of the program; our algorithm, on the other hand, does not require a test suite and localizes the bug using a single trace. 
To compare our algorithm against the existing algorithms, we developed a test generation engine for heap manipulating programs. 
We randomly selected 10 heap manipulating programs from our larger benchmarks (in \tab{tab:all_benchmarks}).
Bugs were injected using our fault injection engine. 
We created 20 versions of the program, each having a single bug for all the selected benchmarks (creating a total of 200 buggy programs). 
For each benchmark, \tab{tab:coverage} shows the average number of test cases generated, and the average statement and branch coverage produced by these test cases averaged over the 20 buggy versions.

To compare across benchmarks of differing sizes, we normalize the average rank of the buggy statement produced by these algorithms with the program size; hence, we report the \textit{developer effort}, i.e., the percentage of the lines of code to be examined before the faulty line is encountered. 

\begin{figure}
        \centering
        \includegraphics[width=0.75\columnwidth]{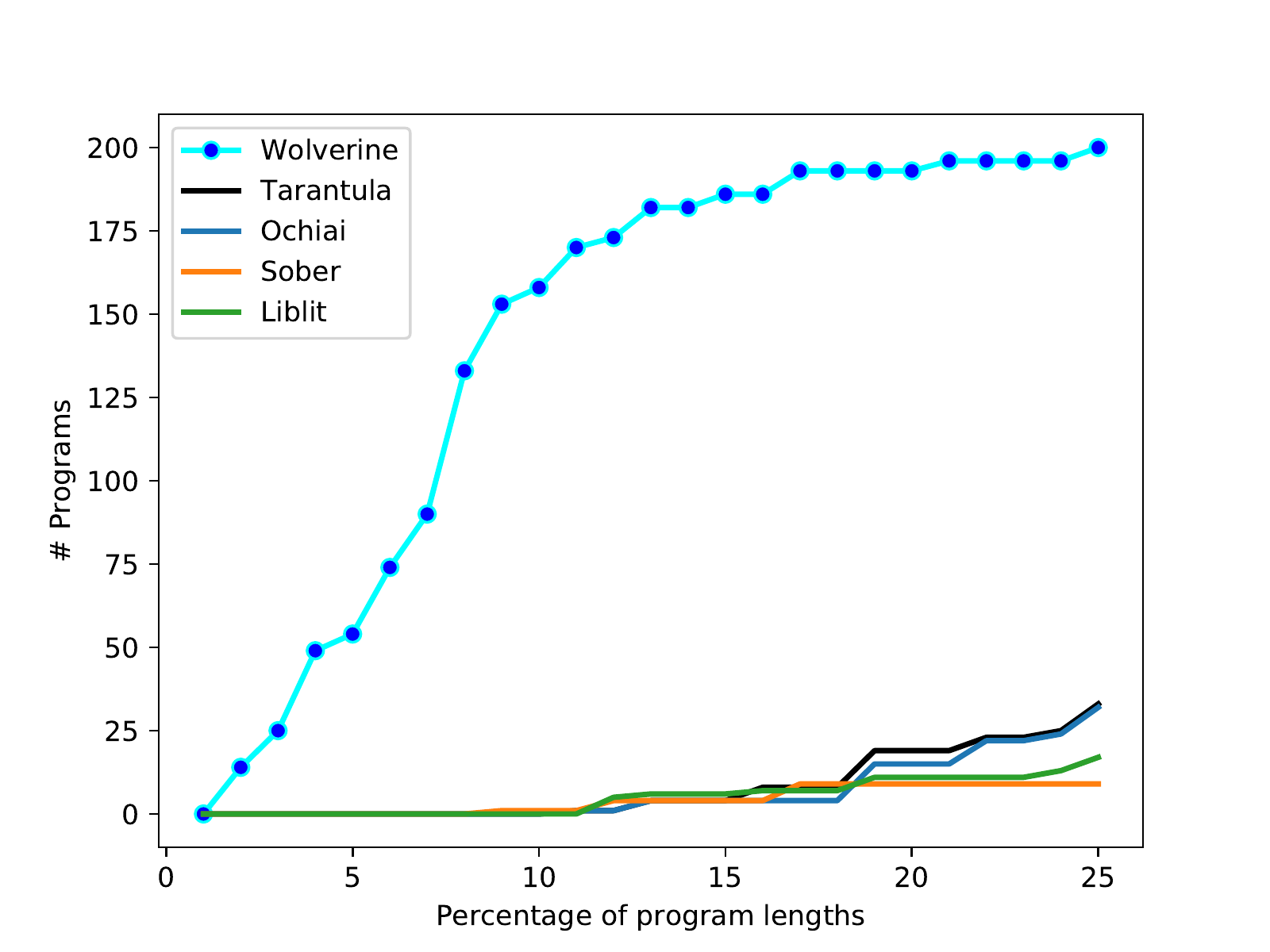}
        \captionof{figure}{Comparison of ranking produced by \toolname and other statistical bug localization algorithms}
        \label{fig:vary_pg_len}
\end{figure}

\fig{fig:vary_pg_len} shows the line plot of the number of programs (out of 200) in which the average rank of the buggy statement produced by different algorithms was within a given effort threshold (\% of program length on the x-axis). 
The plot shows that our algorithm is able to rank the ground truth repair in 150 of the 200 programs to within 10\% of the program size, and all the 200 programs to 25\% of the program size. 
On the other hand, the best performing metric among the other algorithms (Tarantula) is only able to rank 33 of the 200 programs to 25\% of the program size.

\begin{table}
                \centering
                \begin{tabular}{| c | c | c | c |}
                        \toprule
                        Benchmark & \# Tests & Line Coverage & Branch Coverage \\
                        \midrule
                        B1 & 17.0 & 100.0 & 100.0 \\    
                        B2 & 61.0 & 89.0 & 89.0 \\      
                        B3 & 17.0 & 99.0 & 100.0 \\     
                        B4 & 335.0 & 100.0 & 100.0 \\   
                        B5 & 17.0 & 100.0 & 98.0 \\     
                        B6 & 121.0 & 100.0 & 100.0 \\   
                        B7 & 61.0 & 93.0 & 89.0 \\      
                        B8 & 17.0 & 98.0 & 99.0 \\      
                        B9 & 13.0 & 97.0 & 87.0 \\      
                        B10 & 17.0 & 93.0 & 96.0 \\ 
                        \bottomrule
                \end{tabular}
                \caption{Coverage statistics of the tests generated for comparison with other bug localization techniques in ~\fig{fig:vary_pg_len}}
                \label{tab:coverage}
\end{table}

For RQ5, we choose all the 40 heap manipulating programs (in \tab{tab:all_benchmarks}): for each benchmark, $B_i$, at each of the four configurations, we create 40 buggy versions. 
We compare our tool with two configurations of the repair tool (sans localization): 
\begin{enumerate}
 \item \textit{section-wise repair}: when the user has some prior information about the bug and confines the repair tool's search to only a section of the program (like the loop head, a loop body, etc.); a section can contain multiple nested control-flow statements but does not cross loop boundaries;
 \item \textit{unconfined repair}: the repair tool is unleashed on the whole program.
\end{enumerate}

\begin{figure*}[t]
        \centering
        \begin{subfigure}[b]{0.48\textwidth}
                \centering
                \includegraphics[width=1.1\textwidth]{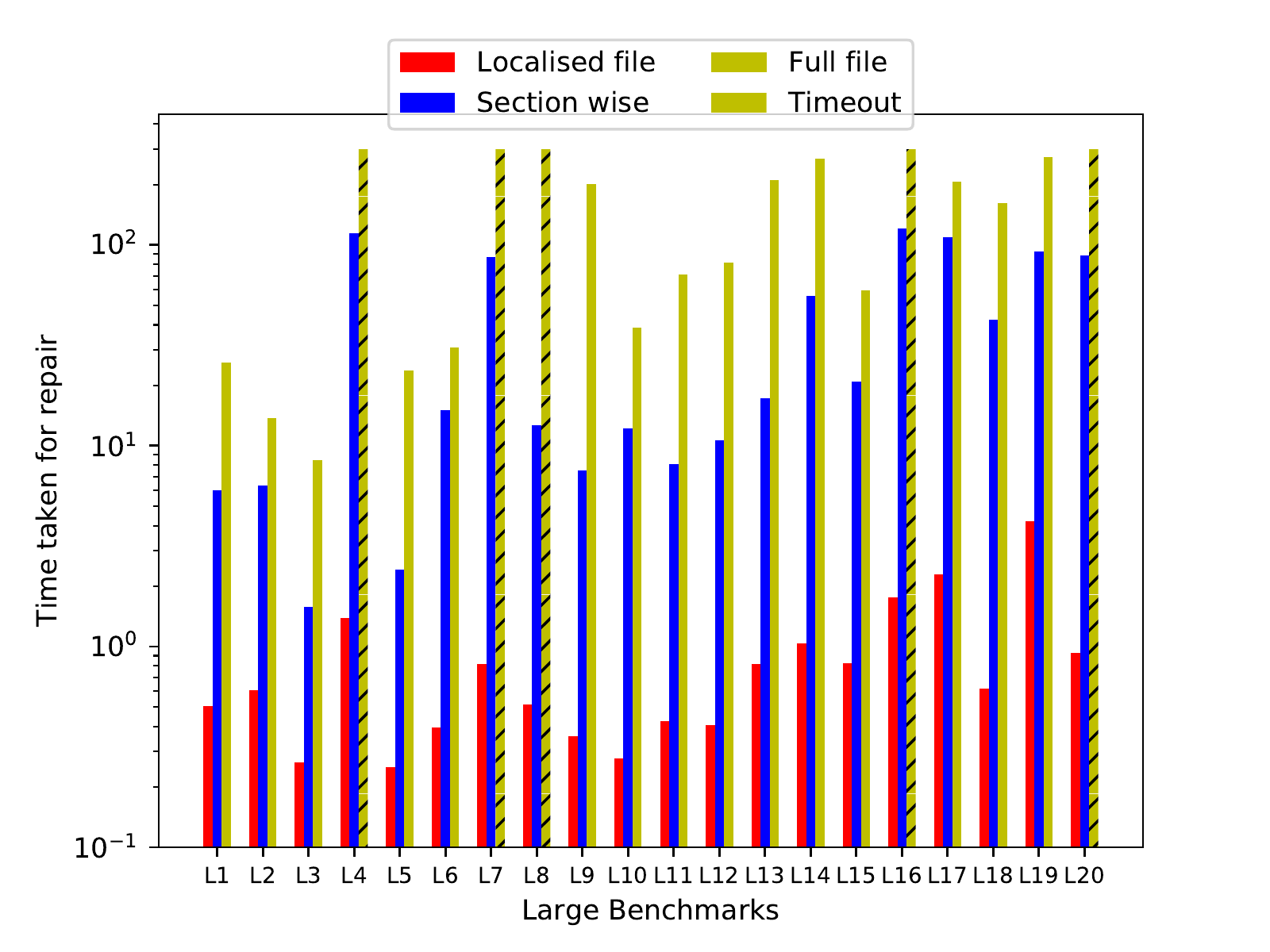}
                \caption{Log scaled timing across localized, sectional-wise and unconfined repair (Class1)}
                \label{fig:large_single_bug}
        \end{subfigure}
		\hfill
        \begin{subfigure}[b]{0.48\textwidth}
                \centering
                \includegraphics[width=1.1\textwidth]{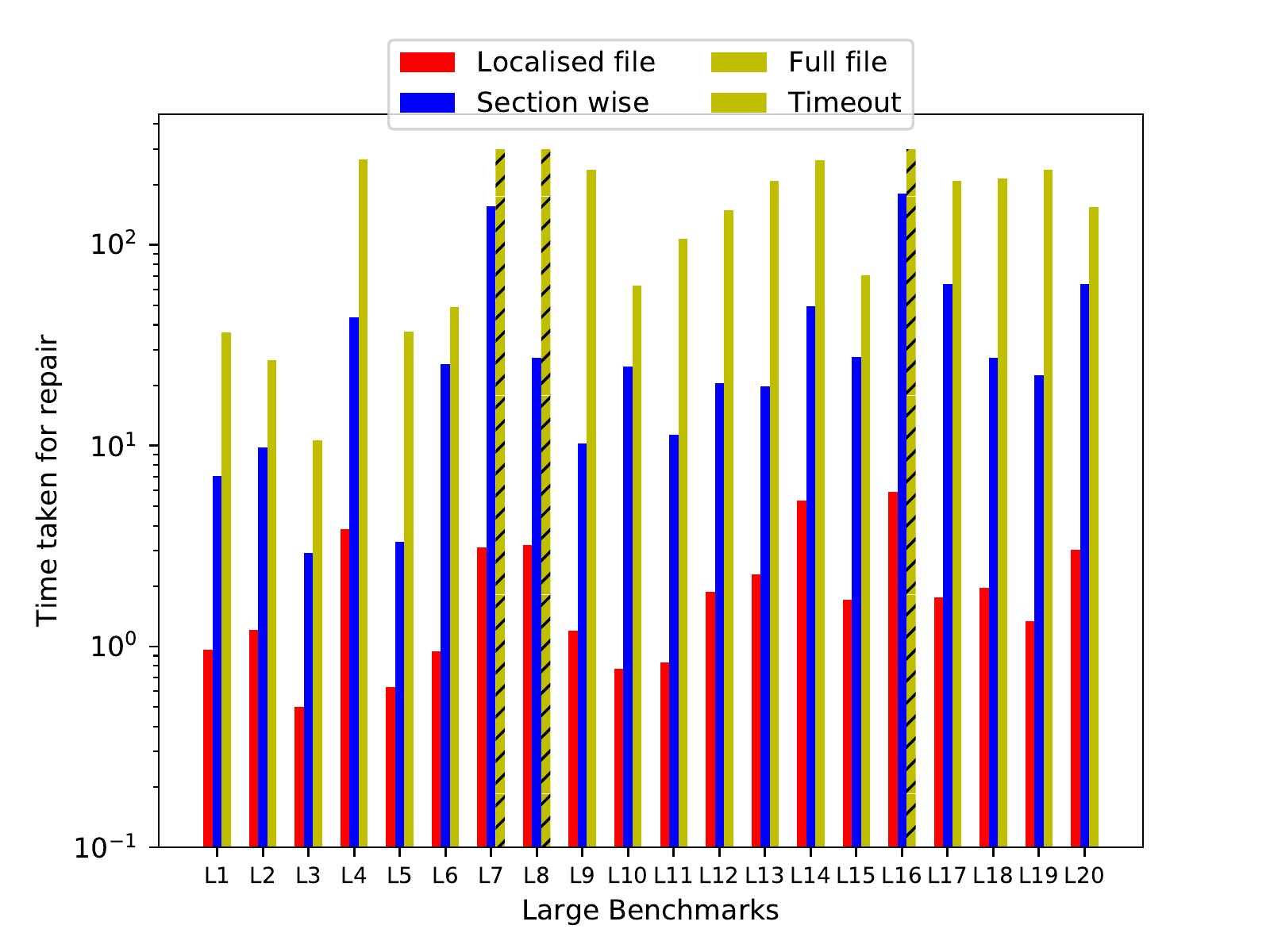}
                \caption{Log scaled timing across localized, sectional-wise and unconfined repair (Class2)}
                \label{fig:large_double_bug}
        \end{subfigure}
        \vspace{10pt}
 		\\
        \begin{subfigure}[b]{0.48\textwidth}
                \centering
                \includegraphics[width=1.1\textwidth]{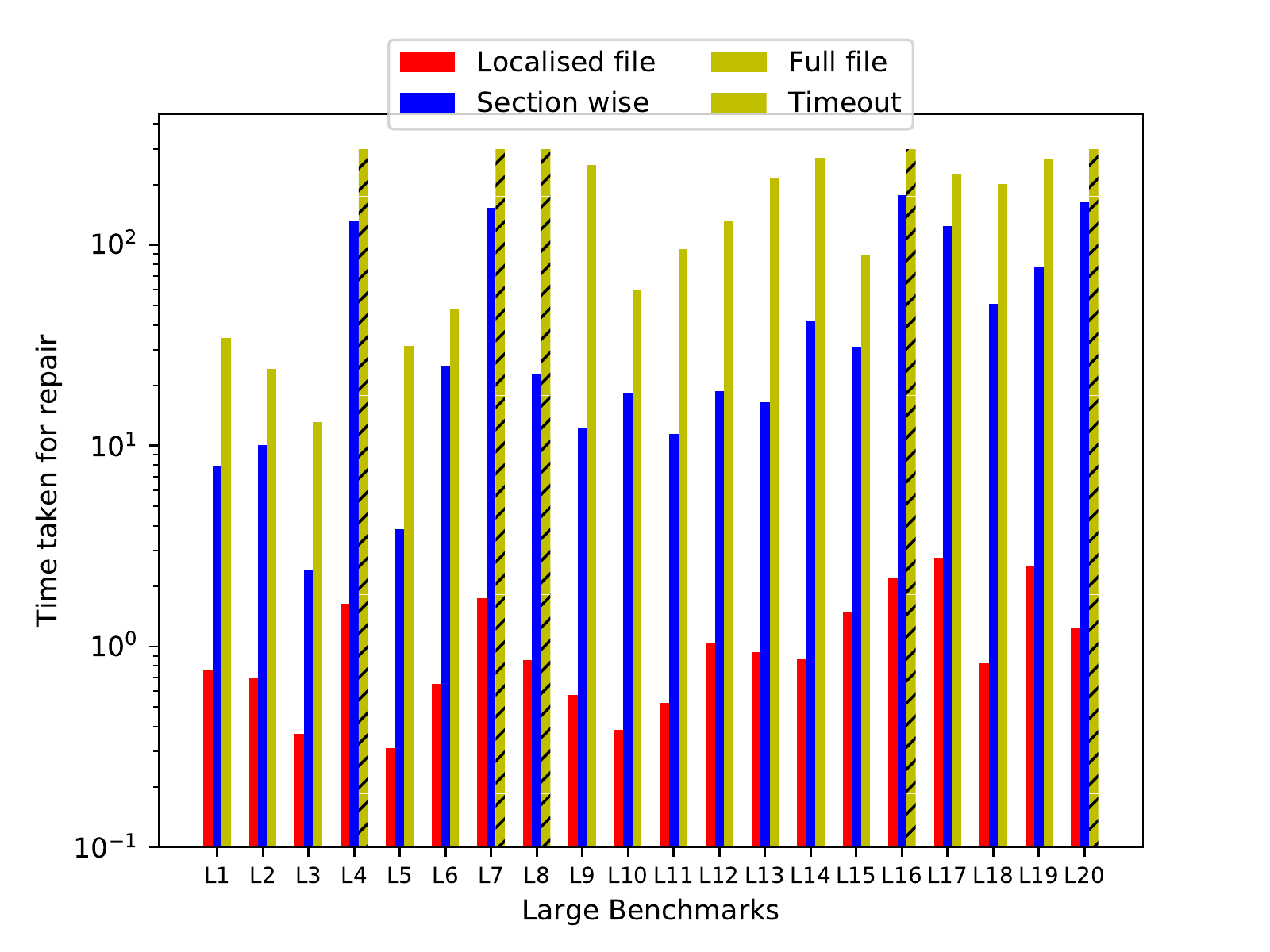}
                \caption{Log scaled timing across localized, sectional-wise and unconfined repair (Class3)}
                \label{fig:large_1bug1slot}
        \end{subfigure}%
        \hfill
        \begin{subfigure}[b]{0.48\textwidth}
                \centering
                \includegraphics[width=1.1\textwidth]{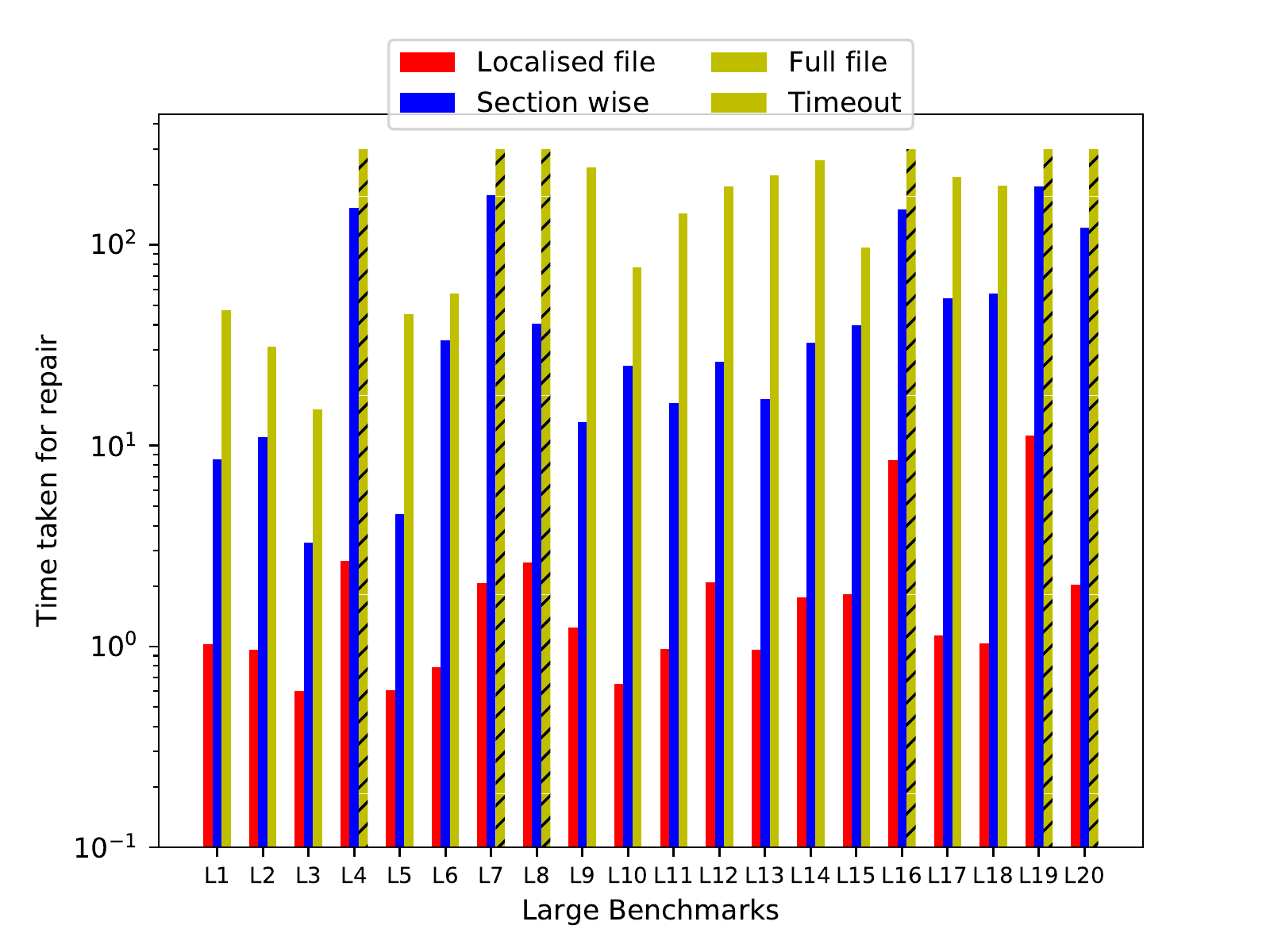}
                \caption{Log scaled timing across localized, sectional-wise and unconfined repair (Class4)}
                \label{fig:large_2bug1slot}
        \end{subfigure}%
    \caption{Time comparison with and without localization (section-wise and unconfined) repair cases (second set of benchmarks)}
    \label{fig:large_with_localization_compare}
\end{figure*}
 
\fig{fig:large_single_bug} compares the repair time (averaged over the 40 buggy versions for single bug configuration) required for \toolname compared in three cases: (1) when localization is used, (2) when the bug is naively localized to a section of the code, and (3) when the bug localization is not used at all. 
The plot is in \textbf{log scale}. It clearly shows that using our bug localization algorithm reduces repair time by several orders of magnitude. 
On average over all benchmarks we were $50\times$ faster compared to the \textit{section-wise repair} and $190\times$ (upto $779\times$ on some benchmarks) faster than in the \textit{unconfined repair} setting. 


\fig{fig:large_double_bug} shows the average repair time compared for the three cases for double bugs (two bugs in the program). 
\fig{fig:large_1bug1slot} shows the average repair time compared for the three cases for single bug and one insert slot configuration. 
\fig{fig:large_2bug1slot} shows the average repair time compared for the three cases for double bug and one insert slot configuration. 

We show a summary of the average speedup results in ~\tab{tab:speedup_results}, both for the \textit{Small} and \textit{Large} instances (we do not provide plots for the smaller instances); understandably, localization benefits the \textit{Large} benchmarks more than the \textit{Small} benchmarks, illustrating the effectiveness of our localization algorithm in the repair of larger, more complex instances. 
For the programs that timed-out, we consider their runtimes as the timeout period (300s).

\begin{table}[t]
        \centering
        \caption{Average speedups gained when compared to case of section-wise repair and case of unconfined repair. 
        The numbers in the parenthesis show the maximum speedup obtained in respective classes for the case of unconfined repair.}
        \begin{tabular}{| c | c | c | c | c |}
        \multicolumn{5}{c}{\textbf{Speed-ups for first set of benchmarks}} \\ \toprule
        Bug Configuration     & Class1      & Class2    & Class3     & Class4    \\ 
        \midrule
        Section-wise repair   &  14         &   4       & 10         & 8          \\ 
        Unconfined repair     &  39 (151)   &   12 (33) & 31 (134)   & 22 (64)    \\
        \bottomrule
        
        \multicolumn{5}{c}{\textbf{Speed-ups for second set of benchmarks}} \\
        \toprule
        Bug Configuration     & Class1      & Class2      & Class3             & Class4     \\ 
        \midrule
        Section-wise repair   &  50         &   20        & 54                 & 28         \\ 
        Unconfined repair     &  190 (779)  &   80 (215)  & 170 (530)          & 86 (257)  \\
        \bottomrule
        \end{tabular}
        \label{tab:speedup_results}
\end{table}

\subsection{Experiments with Student Submissions}
In order to answer RQ6, we collected 247 buggy submissions from students corresponding to 5 programming problems on heap manipulations from an introductory programming course~\cite{Prutor16}. 

\begin{table}
\centering
\caption{Tool evaluation on student submissions}
                                \scalebox{0.9}{
\begin{tabular}{| c || c | c | c | c | c |}
\toprule
Id & Total & Fixed & Impl. Limit & Out of Scope & Vacuous \\
\midrule
P1 & 47 & 30 & 2 & 8 & 7 \\
P2 & 48 & 29 & 3 & 8 & 8 \\
P3 & 48 & 36 & 0 & 5 & 7 \\
P4 & 61 & 46 & 0 & 6 & 9 \\
P5 & 43 & 25 & 0 & 4 & 14 \\
\bottomrule
\end{tabular}
}
\label{tab:student_submission}
\end{table}

We attempted repairing these submissions and categorized a submission into one of the following categories (shown in \tab{tab:student_submission}):


\begin{itemize}
        \item {\bf Fixed}: These are the cases where \toolname could automatically fix the errors.
        \item {\bf Implementation Limitations}: These are cases where, though our algorithm supports these repairs, the current state of our implementation could not support automatic repair.
        \item {\bf Out of scope}: The bug in the submission did not occur in a heap-manipulating statement.
        \item {\bf Vacuous}: In these submissions, the student, had hardly attempted the problem (i.e., the solution is almost empty).
\end{itemize}

Overall, we could automatically repair more than 80\% of the submissions where the student has made some attempt at the problem (i.e., barring the vacuous cases). 

\section{Related Work}
\label{sec:related}

Our proof guided repair algorithm is inspired by a model-checking technique for concurrent programs---referred to as underapproximation widening~\cite{Grumberg:2005}, that builds an underapproximate model of the program being verified by only allowing a specific set of thread interleavings by adding an {\it underapproximation constraint} that inhibit all others. If the verification instance finds a counterexample, a bug is found. If a proof is found which does not rely on the underapproximation constraint, the program is verified; else, it is an indication to relax the underapproximation constraint by allowing some more interleavings. Hence, the algorithm can find a proof from underapproximate models without needing to create abstractions. To the best of our knowledge, ours is the first attempt at adapting this idea for repair. In the case of repairs, performing a proof-guided search allows us to work on smaller underapproximated search spaces that are widened on demand, guided by the proof; at the same time, it allows us to prioritize among multiple repair strategies like insertion, deletion, and mutation. There have also been some attempts at using proof artifacts, like unsat cores, for distributing large verification problems~\cite{Hydra}. In the space of repairs, DirectFix~\cite{DirectFix:2015} also builds a semantic model of a program but instead uses a MAXSAT solver to search for a repair. Invoking a MAXSAT solver is not only expensive, but a MAXSAT solver also does not allow prioritization among repair strategies. In DirectFix, it is not a problem as the tool only allows mutation of a statement for repair and does not insert new statements. Alternatively, one can use a weighted MAXSAT solver to prioritize repair actions, but it is prohibitively expensive; we are not aware of any repair algorithm that uses a weighted MAXSAT solver for repair. 

Inspired by the success of \oldtoolname, there have been proposals at using proof-guided techniques for synthesis and repair: \textsc{Gambit}~\cite{Gambit} uses a proof-guided strategy for debugging concurrent programs under relaxed memory models. It also provides an interactive debugging environment, similar to \toolname, but focusses its debugging/repair attempts at concurrent programs, operating under varying memory models. \textsc{Manthan}~\cite{Manthan1,Manthan2} uses a proof-guided approach to synthesis; instead of starting from a buggy program, its learns an initial version of the program from input-output examples. It, then, uses a similar repair engine as \oldtoolname and \toolname to repair the candidate.

Zimmermann and Zeller~\cite{Zimmermann:2001} introduce {\it memory graphs} to visualize the state of a running program, and Zeller used memory graphs in his popular {\it Delta Debugging} algorithms~\cite{Zeller:2002b,Zeller:2002} to localize faults. Our algorithm is also based on extracting these memory graphs from a concrete execution on \var{gdb} and employing its symbolic form for repair. 
The notion of \texttt{concrete} statement in \toolname bears resemblance to the concolic testing tools~\cite{Godefroid:2005,Sen:2005}. 


Symbolic techniques~\cite{groce:2006,ball:2003,liu:2010,jose:2011,BugAssist:2011,Bavishi:2016,Khurana:2017,Pandey:2019,Gambit} build a symbolic model of a program and use a model-checker or a symbolic execution engine to ``execute" the program; they classify a statement buggy based on the ``distances" of faulty executions from the successful ones. Angelic Debugging~\cite{anagelicDebugging:2011}, instead, uses a symbolic execution engine for fault localization by exploring alternate executions on a set of suspicious locations, while Angelix~\cite{semfix:2013,Angelix:2016} fuses angelic debugging-style fault localization with a component-based synthesis~\cite{Jha:2010} framework to automatically synthesize fixes. There have also been regression aware strategies to localize/repair bugs~\cite{Bavishi:2016}. There have also been proposals to use statistical techniques~\cite{Modi:2013,Sober:2005,liblit:2005}, evolutionary search~\cite{genProg:2012,nguyen:2009,weimer:2009,weimer:2006} and probabilistic models~\cite{prophet:2016} for program debugging. However, though quite effective for arithmetic programs, the above algorithms were not designed for debugging/repairing heap manipulations. There have been proposals that repair the state of a data-structure on-the-fly whenever any consistency check (from a set of checks provided by a user) is found to fail~\cite{Demsky:2003,Juzi:2008}. 
However, our work is directed towards fixing the bug in the source code rather than in the state of the program, which makes this direction of solutions completely unrelated to our problem. In the space of functional programs, there has been a proposal~\cite{Kneuss:2015,Feser:2015} to repair functional programs with unbounded data-types; however, such techniques are not applicable for debugging imperative programs. Finally, \toolname uses a much lightweight technique for fault localization than expensive MAXSAT calls.

There has been some work in the space of synthesizing heap manipulations. The storyboard programming tool~\cite{Singh:2011} uses abstract specifications provided by the user in three-valued logic to synthesize heap manipulations. As many users are averse to writing a formal specification, SYNBAD~\cite{Roy:2013} allows the synthesis of programs from concrete examples; to amplify the user's confidence in the program, it also includes a test-generation strategy on the synthesized program to guide refinement. SYNBAD inspires the intermediate representation of \toolname; \toolname can also be extended with a test-generation strategy to validate the repair on a few more tests before exposing it to the programmer. SYNLIP~\cite{Garg:2015} proposes a linear programming based synthesis strategy for heap manipulations. Feser et al.~\cite{Feser:2015} propose techniques for synthesizing functional programs over recursive data structures. \toolname, on the other hand, attempts repairs; the primary difference between synthesis and repair is that, for a ``good" repair, the tool must ensure that the suggested repair only makes ``small" changes to the input program rather than providing a completely alternate solution. Other than synthesis of heap manipulations, program synthesis has seen success in many applications, from bit-manipulating programs~\cite{Jha:2010}, bug synthesis~\cite{BugSynthsis}, parser synthesis~\cite{Leung:2015,Singal:2018} and even differentially private mechanisms~\cite{Kolahal}.
Fault localization techniques have seen both statistical and formal algorithms. Statistical debugging techniques~\cite{Sober:2005,liblit:2005,Tarantula:2005,Oichai:2009,Ulysis,Modi:2013} have been highly popular for large code-bases. These techniques essentially attempt to discover correlations between executions of parts of the program and its failure. However, though these techniques work quite well for large codebases, they are not suitable for somewhat smaller, but tricky programs, like heap manipulations. Our experiments (\textbf{RQ4}) demonstrate this and thus motivate different fault localization techniques for such applications.
Moreover, these techniques essentially provide a ranking of the suspicious locations and hence are somewhat difficult to adopt with repair techniques. On the other hand, our localization algorithm provides a sound reduction in the repair space, thereby fitting quite naturally with the repair.



\section{Discussion and Conclusion}
\label{sec:discussion}
We believe that tighter integration of dynamic analysis (enabled by a debugger) and static analysis (via symbolic techniques) can open new avenues for debugging tools. 
This work demonstrates that a concrete execution on a debugger to collect the potentially buggy execution and the user-intuitions on the desired fixes, fed to a bug localizer that contracts the repair space, and a proof-directed repair algorithm on the reduced search space, is capable of synthesizing repairs on non-trivial programs in a complex domain of heap-manipulations. 
We are interested in investigating more in this direction. 
 
There exist threats to validity to our experimental results, in particular from the choice of the buggy programs and how the bugs were injected. 
We were careful to select a variety of data-structures and injected bugs via an automated fault injection engine to eliminate human bias; nevertheless, more extensive experiments can be conducted.


\bibliographystyle{plain}
\bibliography{refs} 

\noindent \Large{\bf Appendix}
\appendix

\section{Theoretical Analysis of the Bug Localization Algorithm}
  Let $\mathcal{I}$ be the identity function that copies the $\omega_{pre} \text{ or } \omega_{post}$. Then,

    \noindent $[\omega^f_0, \omega^f_1, \omega^f_2, \dots, \omega^f_n] \gets [\mathcal{I}, \Delta_{1, 1}, \Delta_{1, 2}, \dots \Delta_{1, n}] (\omega_{pre}$, P) \\
    $[\Omega^b_0, \Omega^b_1, \Omega^b_2, \dots, \Omega^b_n] \gets [\nabla_{n, 1}, \nabla_{n, 2}, \dots \nabla_{n, n}, \mathcal{I}] (\omega_{post}$, P) \\
    $[\Gamma_0, \Gamma_1, \Gamma_2, \dots, \Gamma_n] \gets [\omega^f_0, \omega^f_1, \omega^f_2, \dots, \omega^f_n] \otimes [\Omega^b_0, \Omega^b_1, \Omega^b_2, \dots, \Omega^b_n]$; \hfill $\Omega$ is a set of states while $\omega$ is a single state. \\

\begin{figure*}
        \centering
        \begin{subfigure}[b]{\textwidth}
                \centering
                \includegraphics[width=\textwidth]{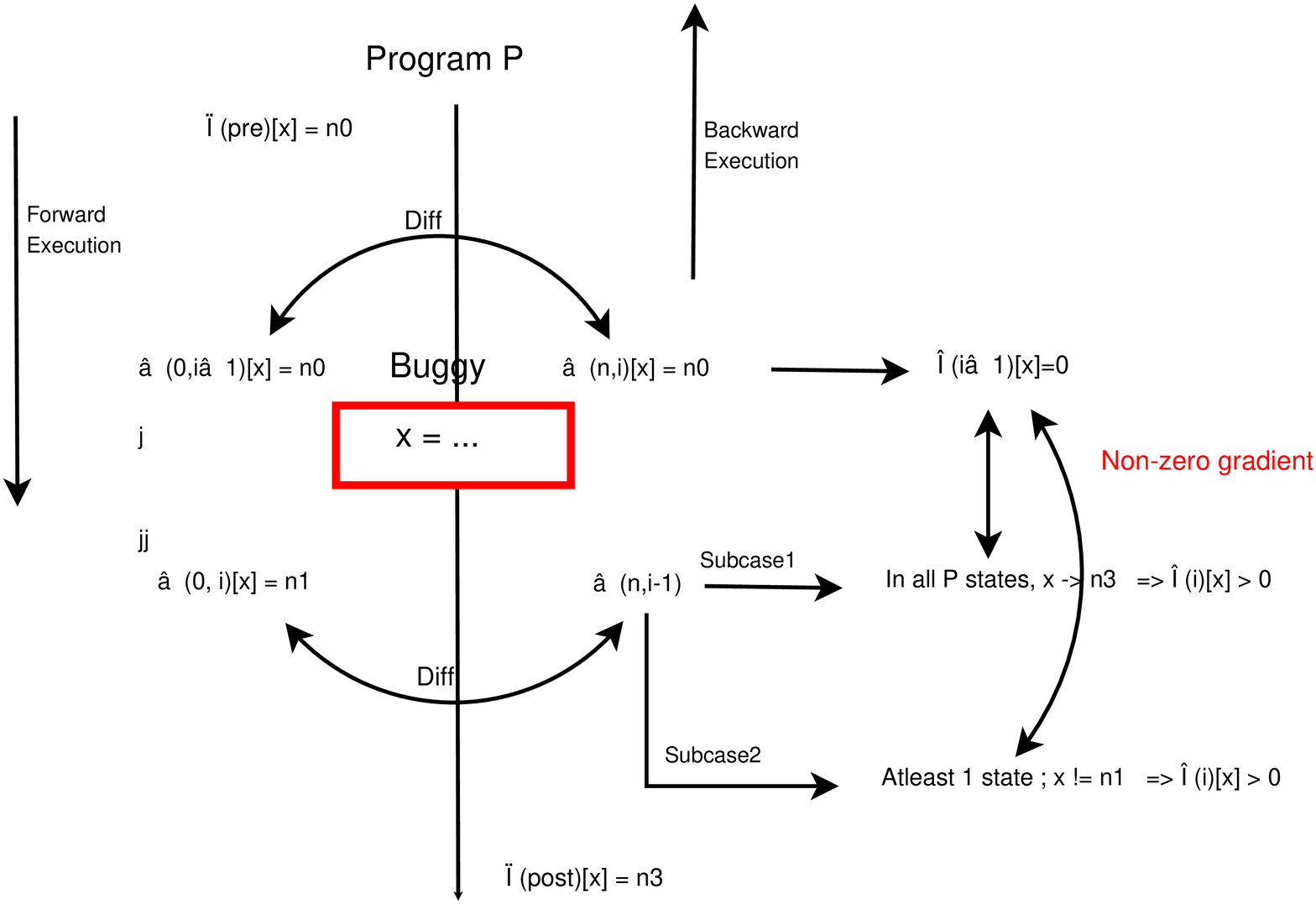}
                \caption{Illustrations for case 1}
                \label{fig:proof_case1}
        \end{subfigure}%
        \\ \vspace{4em}
        \begin{subfigure}[b]{\textwidth}
                \centering
                \includegraphics[width=\textwidth]{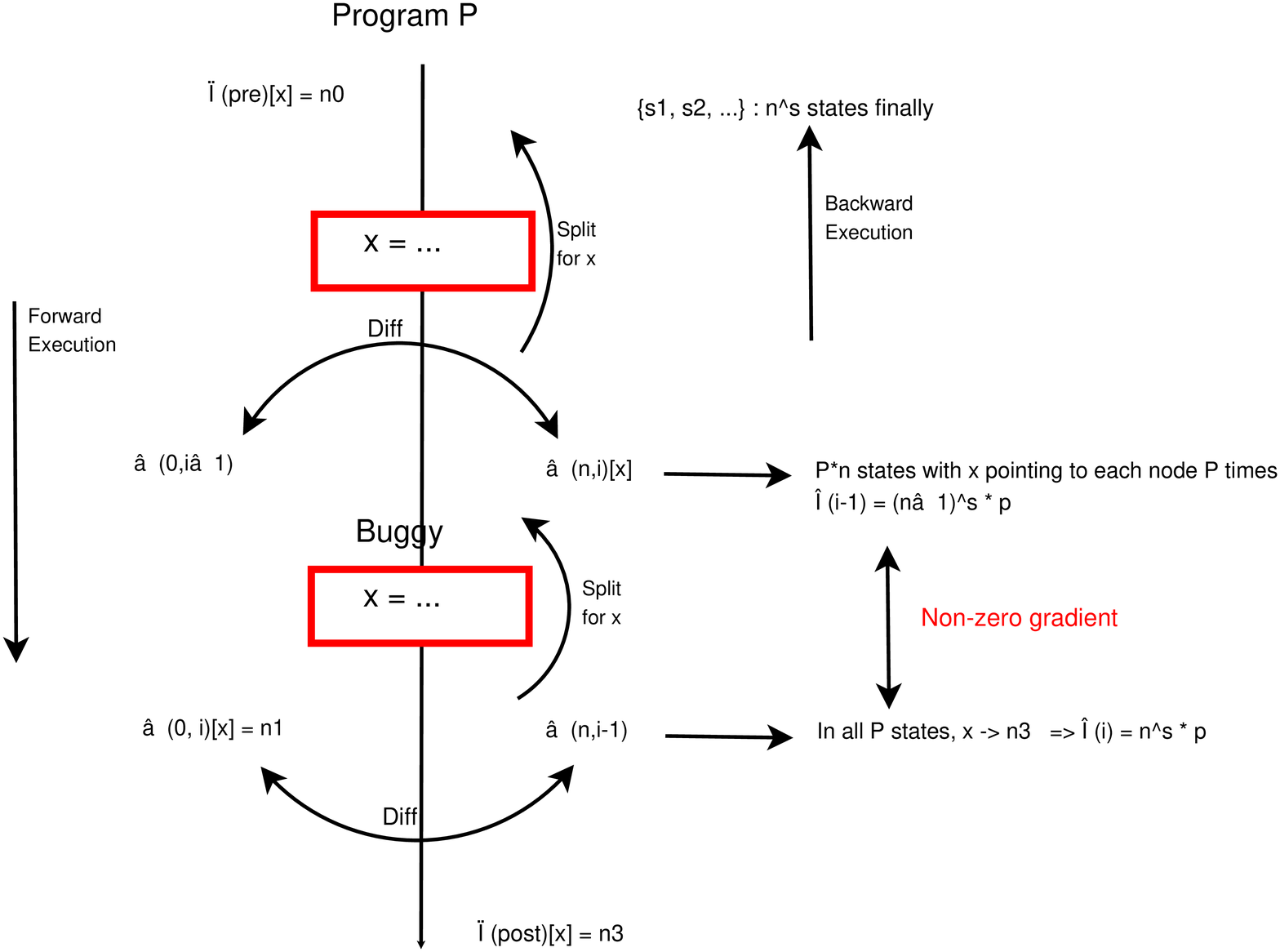}
                \caption{Illustrations for case 3}
                \label{fig:proof_case3}
        \end{subfigure}
        \caption{Illustrations for the proof of theorem~\ref{theorem_bug_local1}}
        \label{fig:illustration_proof}
\end{figure*}

\begin{theorem}
    \textbf{If a real bug is at location $i$, it implies that $i$ is in the suspicious set}
\label{theorem_bug_local1}
\end{theorem}

\begin{proof}

Let $\hat{P}$ and $P$ denote the correct and buggy programs, respectively.

\textbf{Assumption:} Assume that in the trace $P_n$ (sequence of statements of length n), the $i^{th}$ statement has a real bug and $LHS(s_i)$ be $x$. 
Let us define the correct execution as the forward execution states of $\hat{P}$ at each program point.
Let $\Delta_{0, i}(\omega_{pre}, \hat{P})[x] = \var{n1}$, $\Delta_{0, i}(\omega_{pre}, P)[x] = \var{n2}$, $\omega_{post}[x] = \var{n3}$ and $\omega_{pre}[x] = \var{n0}$. Therefore, the number of nodes and points-to pairs in the datastructure, $\var{n} \geq 2 $.  \hfill \text{[Assump1]}

\begin{case}
        The $i^{th}$ statement is upward exposed: this is illustrated in \fig{fig:proof_case1}.
        \begin{enumerate}
        \item $\Delta_{0, i-1}(\omega_{pre}, P)$ = $\Delta_{0, i-1}(\omega_{pre}, \hat{P})$  \hfill [Since the bug is in $i^{th}$ stmt]
        \item $\nabla_{n, i}(\omega_{post}, P)[x] = \omega_{pre}[x]$  \hfill [$asgn_{1}$ rule in \fig{fig:back_semantics}]
        \item $\Delta_{0, i-1}(\omega_{pre}, P)[x] = \omega_{pre}[x]$  \hfill [Lemma ~\ref{lemma_basic1}]
        \item $\nabla_{n, i}(\omega_{post}, P)[x] = \Delta_{0, i-1}(\omega_{pre}, P)[x]$ \hfill [From 2 and 3]
        \item $\Gamma_{i-1}[x] = 0$ \hfill [Since $\Gamma_{i-1} = (\nabla_{n, i} - \Delta_{0, i-1}) * (multiplying factor)$ and $\nabla_{n, i} - \Delta_{0, i-1} = 0$]
        \item Also since $i^{th}$ statement is upward exposed there is no previous splitting of states when this statement is backtracked. \hfill [Backward Semantics in \fig{fig:back_semantics}]
        \item Without loss of generality, lets say there are already \var{p} backtracked states before $i^{th}$ statement is backtracked. 
        \item Subcase 1: There has not been any splitting for $x$ in the backward execution. Therefore for all the \var{p} states (in $\nabla_{n, i+1}$), $x$ points-to $\var{n3}$ \hfill [Stmt is both upward and downward exposed]
        \item $\Gamma_{i}[x] = p * multiplying factor > 0$ \hfill [From 8 and since $\Gamma_{i} = (\nabla_{n, i+1} - \Delta_{0, i}) * (multiplying factor)$]
        \item Subcase 2: There has been a previous splitting for $x$ in the backward execution. Therefore in atleast one of the \var{p} states, the value of $x \neq \var{n1}$. This is because splitting happens uniformly across all nodes and there are atleast 2 nodes in the data-structure. \hfill [Rules in \fig{fig:back_semantics}]
        \item $\Gamma_{i}[x] > 0$ \hfill [From 10]
        \item Hence for loop at step 5 in \alg{alg:buglocalization}, since $\Gamma_{i-1} \neq \Gamma_{i}, i^{th}$ statement will be added to suspicious set.
        \end{enumerate}
\end{case}

\begin{case}
       The $i^{th}$ statement is sandwiched. We add all sandwiched statements to the suspicious set since the program loses all information regarding the correctness of such a statement.

\end{case}

\begin{case}
       The $i^{th}$ statement is downward exposed and not upward exposed: this is illustrated in \fig{fig:proof_case3}.
        \begin{enumerate}
        \item  There will be splitting of states when $i^{th}$ statement is backtracked.  \hfill [$asgn_{2}$ rule in \fig{fig:back_semantics}]
        \item  Without loss of generality, lets say there are \var{p} backtracked states before $i^{th}$ statement is backward executed.
        \item  In all these \var{p} states, the value of variable $x = \var{n3}$.      \hfill [Lemma ~\ref{lemma_basic2}]
        \item Let there be $\var{s}$ splits for $x$ until the full backtrack of the program.
        \item The final number of states after backtracking is $n^{s} = M$, where $n$ is the number of nodes and points-to pairs \hfill [Rules in \fig{fig:back_semantics}]
        \item Each of the p states contribute a difference of 1 in $\Gamma_{i} = p * \frac{n^s}{p} = n^s$ \hfill [$\Gamma_{i} = (\nabla_{n, i+1} - \Delta_{0, i}) * (multiplying factor)$]
        \item $\Gamma_{i-1} = \frac{n^s}{p*n} * (n-1)*p = n^s - n^{s-1}$  \hfill [Out of n nodes, 1 will match with a node in $\Delta_{0, i}(\omega_{pre}, {P})$; uniform splitting]
        \item $\Gamma_{i-1} - \Gamma_{i} = n^s - n^{s-1} - n^s \ne 0$, for $n > 1$   \hfill [Since number of nodes and points-to pairs > 1]
        \item Hence, for the loop at step 5 in \alg{alg:buglocalization}, since $\Gamma_{i-1} \neq \Gamma_{i}, i^{th}$ statement will be added to suspicious set.
        \end{enumerate}
\end{case}

In all cases, \alg{alg:buglocalization} successfully catches the buggy statement.











\end{proof}


\begin{theorem}
\textbf{If $C$ is the set of suspicious locations captured by \alg{alg:buglocalization}, then replacing all statements $s \notin C$ by $lhs = *$ maintains the ground truth repair in the proof-guided repair algorithm.}
\end{theorem}
\begin{proof}
        Any statement which depends on a buggy statement is also put in the suspicious set by our algorithm. 
        This is natural because upon execution of the buggy statement, an incorrect state is achieved, and any statements which further depend on this state would be buggy with respect to the ground truth state. 
        Hence, we can slice away the statement from repair by making them non-deterministic.
\end{proof}


\end{document}